\documentclass[useAMS,usedcolumn,usegraphicx,usenatbib]{mn2e}
\usepackage{amssymb,amstext,amsfonts} 
\usepackage[fleqn]{amsmath}
\usepackage{subfigure}
\usepackage{dcolumn}
\usepackage{hyperref}

\newcommand{\eqnref}[1]{(\ref{eq:#1})}
\newcommand{\figref}[1]{fig.~\ref{fig:#1}}
\newcommand{\Figref}[1]{Figure~\ref{fig:#1}}
\newcommand{\tabref}[1]{table~\ref{tab:#1}}
\newcommand{\secref}[1]{Sec.~\ref{sec:#1}}

\newcommand{\apref}[1]{Appendix~\ref{sec:#1}}

\DeclareMathOperator{\Ei}{Ei}
\DeclareMathOperator{\erf}{erf}
\DeclareMathOperator{\Beta}{B}

\newcommand{\units}[1]{\ensuremath{~\mathrm{#1}}}

\newcommand{\sub}[1]{\ensuremath{_\mathrm{#1}}}
\newcommand{\super}[1]{\ensuremath{^\mathrm{#1}}}
\newcommand{\dd}{\ensuremath{\mathrm{d}}}
\newcommand{\diff}[2]{\ensuremath{\dfrac{\dd {#1}}{\dd {#2}}}}

\newcommand{\intd}[4]{\ensuremath{\displaystyle \int_{#1}^{#2}{#3}\,\dd{#4}}}
\newcommand{\recip}[1]{\ensuremath{\dfrac{1}{#1}}}

\newcommand{\order}[1]{\ensuremath{\mathcal{O}({#1})}}

\newcommand{\innerprod}[2]{\ensuremath{\left({#1}\middle|{#2}\right)}}

\title[Expectations for EMRBs from the GC]{Expectations for extreme-mass-ratio bursts from the Galactic Centre}
\author[C.\ P.\ L.\ Berry and J.\ R.\ Gair]{C.\ P.\ L.\ Berry$^{1}$\thanks{E-mail:
cplb2@cam.ac.uk} and J.\ R.\ Gair$^{1}$\\
$^{1}$Institute of Astronomy, University of Cambridge, Madingley Road, Cambridge, CB3 0HA}

\begin{document}

\date{\today}

\pagerange{\pageref{firstpage}--\pageref{lastpage}} \pubyear{2013}

\maketitle

\label{firstpage}

\begin{abstract}
When a compact object on a highly eccentric orbit about a much more massive body passes through periapsis it emits a short gravitational wave signal known as an extreme-mass-ratio burst (EMRB). We consider stellar mass objects orbiting the massive black hole (MBH) found in the Galactic Centre. EMRBs provide a novel means of extracting information about the MBH; an EMRB from the Galactic MBH could be highly informative regarding the MBH's mass and spin if the orbital periapsis is small enough. However, to be a useful astronomical tool EMRBs must be both informative and sufficiently common to be detectable with a space-based interferometer. We construct a simple model to predict the event rate for Galactic EMRBs. We estimate there could be on average $\sim 2$ bursts in a two-year mission lifetime for {\textit{LISA}}. Stellar mass black holes dominate the event rate. Creating a sample of $100$ mission realisations, we calculate what we could learn about the MBH. On average, we expect to be able to determine the MBH mass to $\sim1\%$ and the spin to $\sim0.1$ using EMRBs.
\end{abstract}


\section{Introduction}

The most compelling evidence for the existence of astrophysical black holes (BHs) comes from the measurement of stellar orbits at the centre of the Galaxy. The stars are found to orbit an object of mass $M_\bullet \simeq 4 \times 10^6 M_\odot$ coincident with the compact radio source Sagittarius A* \citep{Reid2004, Ghez2008, Gillessen2009, Meyer2012}. This is the nearest member of a population of massive black holes (MBHs; \citealt{Volonteri2010}) that are believed to occupy the centres of galaxies \citep{Lynden-Bell1969, Rees1984, Ferrarese2005}. The Galactic Centre (GC) is an ideal laboratory for investigating the properties of an MBH and its surrounding nuclear star cluster \citep{Genzel2010}.

One means of investigating the properties of MBHs is through gravitational waves (GWs). A stellar mass compact object (CO), such as a main sequence (MS) star, white dwarf (WD), neutron star (NS) or stellar mass BH, emits gravitational radiation as it orbits the MBH. A space-borne detector, such as the \textit{Laser Interferometer Space Antenna} (\textit{LISA}) or the \textit{evolved Laser Interferometer Space Antenna} (\textit{eLISA}), is designed to be able to detect GWs in the frequency range of interest for these encounters assuming a MBH mass of $\sim 10^4$--$10^7M_\odot$ \citep{Bender1998, Danzmann2003, Jennrich2011, Amaro-Seoane2012a}. There are currently no funded space-borne detector missions. However, the European Space Agency's \textit{LISA Pathfinder} will be launched in 2015 and demonstrate the key technologies required for a successful space-borne mission \citep{Anza2005, Antonucci2012}. Following on from previous work, we use the classic \textit{LISA} design; this should provide a sensible benchmark for any future detectors.

The gravitational waveforms emitted from extreme-mass-ratio systems have been much studied \citep{Glampedakis2005, Barack2009}. On account of the extreme-mass-ratio between the two bodies, we can approximate the CO as moving in the background spacetime of the MBH. The GWs carry away energy and angular momentum, causing the orbit to shrink until eventually the object plunges into the MBH. The primary focus has been upon the later stages of the orbital evolution, the last $1$--$2\units{yr}$ immediately preceding the plunge. By this point, the orbit has nearly circularised and emits continuously within the detector's frequency band. These signals are extreme mass-ratio inspirals (EMRIs; \citealt{Amaro-Seoane2007}). EMRIs can be observed over many orbits, allowing exquisitely high signal-to-noise ratios (SNRs) to accumulate. This makes them excellent probes of the background geometry permitting precise measurements of the system parameters and tests of general relativity.

EMRIs evolve from more eccentric orbits. These initial orbits may be the results of scattering from two-body encounters. Rather than emitting a continuously detectable signal, highly eccentric orbits only emit significant radiation in a burst around the point of closest approach to the MBH. These are extreme mass-ratio bursts (EMRBs; \citealt*{Rubbo2006}).

EMRBs are much shorter in duration than EMRIs. This means they do not accumulate as high SNRs or produce as detailed maps of the spacetime. They are therefore less valued prizes. However, they may still be an interesting signal. As an object inspirals, it emits many bursts before eventually settling into a low-eccentricity EMRI. Some objects shall be scattered by two-body encounters and never reach the EMRI phase \citep{Alexander2003}. Thus, there are many potential EMRBs per EMRI, although this does not necessarily translate to there being more detectable EMRBs than EMRIs.

For EMRBs to be a useful astronomical signal, we require three things: the bursts contain sufficient information to improve our knowledge of their source systems, their event rate is sufficiently high that we expect to observe them over a mission lifetime and the signals can be successfully extracted from the data stream.

We have previously addressed the first requirement: EMRBs can give good constraints on the key parameters describing the Galaxy's MBH if the periapse distance is $r\sub{p} \lesssim 10 r\sub{g}$, where $r\sub{g} = GM_\bullet/c^2$ is a gravitational radius \citep{Berry2013}. This would allow us to improve upon the current uncertainty in the mass measurement of $8\%$ \citep{Gillessen2009}. In addition, we could also measure the spin magnitude to a precision of better than $0.1$.

The second requirement shall be the subject of this work. Previously, the best estimate for the event rate was given by \citet*{Hopman2007}; they predicted the event rate for \textit{LISA} is $\sim 1\units{yr^{-1}}$. We follow a similar approach, but, significantly, we improve the calculation of SNR by using numerical kludge (NK) waveforms \citep{Babak2007}. Our models differ in a number of small ways; most notably, we allow for bursts to come from objects in the earliest stages of GW inspiral. In addition to this, we extend the analysis by not only considering the number of events that would be detectable, but also how many would be informative.

We begin in \secref{Waveforms} by recapping how to generate and analyse burst waveforms. In \secref{Rates}, we present our model for calculating event rates and discuss its ingredients. Whilst we have been careful in trying to account for all the relevant effects, the model remains approximate and can only be trusted to give order-of-magnitude estimates. Our results, including the predicted event rate along with a quantification of the estimated information content, are presented in \secref{Gal-Results} and discussed further in our conclusion, \secref{Gal-End}.

\section{Waveforms and parameter uncertainties}\label{sec:Waveforms}

To establish the detectability and usefulness of EMRBs, it is necessary to calculate model waveforms. This is done using the numerical kludge approximation. We use exactly the same construction as is described in \citet{Berry2013}; the key details are outlined below.

The detectability of a burst is dependent upon its SNR. To save calculating SNRs directly, it is possible to estimate the value from the periapse radius using a simple scaling relation. This is introduced in \secref{SNR}. We make use of this when selecting orbits of potential interest, but calculate the SNR from waveforms for more accurate results.

Once we have determined which bursts are of interest, we can evaluate the accuracy to which parameters can be determined, should that burst be observed. To do so, we perform Markov chain Monte Carlo (MCMC) simulations \citep[chapter 29]{MacKay2003}.

\subsection{Numerical kludge waveforms}\label{sec:NK}

Extreme-mass-ratio signals can be simulated in a computationally efficient manner using a semirelativistic approximation \citep{Ruffini1981}: we assume that the CO moves along a geodesic in the Kerr geometry, but radiates as if it were in flat spacetime. This technique is known as an NK. The justification of this technique is through comparison with more accurate, and computationally intensive, methods \citep{Gair2005, Babak2007}. Using a geodesic for the trajectory should ensure that the signal has the correct frequency components but neglecting the effects of background curvature means that these need not have the correct amplitudes. The typical errors in the amplitude profile of a waveform can be a few percent \citep{Tanaka1993,Gair2005}: the total amplitude error in our waveforms, integrating over all frequencies, has been estimated as typically $\sim 5\%$, increasing to $\sim 10\%$ for the most relativistic orbits ($r\sub{p} \lesssim 4 r\sub{g}$; \citealt{Berry2013}). We only consider parabolic (marginally bound) orbits, where the orbiting body would be at rest at infinity. When calculating event rates we consider a range of highly eccentric orbits, not just those that are exactly parabolic. This introduces an additional error into the waveforms, but simplifies the analysis. The modification to the burst waveform should be small, as the trajectories are close to being parabolic; the largest differences are at large radii, which are unimportant for the waveforms.

To construct our NK waveforms, we first integrate the Kerr geodesic equations of motion. To avoid difficulties with turning points in the trajectory, we employ angular variables in place of the radial and polar coordinates \citep{Drasco2004}
\begin{align}
r = {} & \dfrac{2 r\sub{p}}{1 + \cos\psi};\\
\cos^2\theta = {} & \dfrac{Q}{Q+L_z^2}\cos^2\chi,
\end{align}
where $Q$ is the Carter constant and $L_z$ is the angular momentum about the $z$-axis.

Once the Kerr geodesic is constructed, we identify the Boyer--Lindquist coordinates with flat-space spherical polars \citep{Gair2005, Babak2007}. Using the relativistic trajectory ensures that the waveform incorporates the correct frequency components; translating to flat space means we can make use of the flat-space wave-generation formula. The downside of this is that the waveform amplitude shall not be precisely correct.

We use the quadrupole--octupole formula for the gravitational strain \citep{Bekenstein1973, Press1977, Yunes2008}. This is the familiar quadrupole formula (\citealt*[section 36.10]{Misner1973}; \citealt[section 17.9]{Hobson2006}), plus the next order terms. The higher order terms modify the amplitudes of some frequency components by a few tens of percent for the more relativistic orbits, although the integrated effect is smaller.

\subsection{SNR scaling}\label{sec:SNR}

The SNR $\rho$ of a particular burst depends upon the precise shape of its trajectory (as specified by initial conditions) and the position of the detector. However, the most important parameter is the periapse distance.

The $\rho$--$r\sub{p}$ relation is largely determined by the shape of the noise curve. For our simulations, we employ the noise model of \citet{Barack2004}. For bursts from the GC, over much of the range of interest, the curve can be approximated as a simple power law \citep{Berry2013}
\begin{equation}
\log\rho \simeq -2.7\log\left(\dfrac{r\sub{p}}{r\sub{g}}\right) + \log\left(\dfrac{\mu}{M_\odot}\right) + 4.9,
\label{eq:SNR-power-law}
\end{equation}
where $\mu$ is the mass of the CO.

We assume a detection threshold of $\rho = 10$. This gives expected detection limits on the periapse radius. With a $1 M_\odot$ CO, bursts should be detectable for $r\sub{p} \lesssim 27 r\sub{g}$ and with a $10 M_\odot$ CO for $r\sub{p} < 65 r\sub{g}$.

\subsection{Parameter estimation}\label{sec:param-est-MCMC}

Once we have a detected signal $\boldsymbol{s}(t)$, we can consider the inference of the source parameters $\boldsymbol{\lambda}$. The waveform depends on the properties of the MBH; the CO and its orbit, and the detector.

We assume the position of the detector is known, and the MBH is coincident with the radio source of Sgr A* which is thought to be within $20 r_\mathrm{g}$ of the MBH \citep{Reid2003,Doeleman2008}. We use the J2000.0 coordinates which are determined to high accuracy \citep{Reid1999, Yusef-Zadeh1999}.

The parameters left to infer are \citep{Berry2013}:
\begin{enumerate}
\item[(1)] The MBH's mass $M_\bullet$. This is well constrained by the observation of stellar orbits about Sgr A* \citep{Ghez2008, Gillessen2009}; we employ the estimate $M_\bullet = (4.31 \pm 0.36) \times 10^6 M_\odot$ whenever a mass is required.
\item[(2)] The spin parameter $a_\ast$. This is constrained to the range $|a_\ast| < 1$.
\item[(3, 4)] The orientation angles for the BH spin $\Theta_\mathrm{K}$ and $\Phi_\mathrm{K}$. These are measured relative to the polar axis of the ecliptic coordinate system commonly used for describing the positions of celestial objects.
\item[(5)] The ratio of the GC distance and the CO mass $\zeta = R_0/\mu$. This scales the amplitude of the waveform. The distance has been measured using stellar orbits to be $R_0 = 8.33 \pm 0.35~\mathrm{kpc}$ \citep{Gillessen2009}.
\item[(6, 7)] The angular momentum of the CO, parametrized by the magnitude at infinity $L_\infty = \sqrt{L_z^2 + Q}$ and the orbital inclination $\iota = \tan^{-1}(\sqrt{Q}/L_z)$.
\item[(8--10)] Coordinates specifying the trajectory. We use the angular phases at periapse, $\phi_\mathrm{p}$ and $\chi_\mathrm{p}$, as well as the time of periapse $t_\mathrm{p}$.
\end{enumerate}
The first four are specific to the MBH, and we shall attempt to quantify the constraints we can expect to place on these from EMRBs.

The probability that the source is described by parameters $\boldsymbol{\lambda}$ is given by the posterior distribution
\begin{equation}
p(\boldsymbol{\lambda}|\boldsymbol{s}(t)) = \dfrac{p(\boldsymbol{s}(t)|\boldsymbol{\lambda})p(\boldsymbol{\lambda})}{p(\boldsymbol{s}(t))}.
\end{equation}
Here $p(\boldsymbol{s}(t)|\boldsymbol{\lambda})$ is the likelihood of the parameters, $p(\boldsymbol{\lambda})$ is the prior probability distribution for the parameters, and the evidence $p(\boldsymbol{s}(t))$ is a normalisation factor.

If the parameter set $\boldsymbol{\lambda}_0$ defines a waveform $\boldsymbol{h}_0(t) = \boldsymbol{h}(t; \boldsymbol{\lambda}_0)$, the likelihood of the parameters is
\begin{equation}
p(\boldsymbol{s}(t)|\boldsymbol{\lambda}_0) \propto \exp\left[-\recip{2}\innerprod{\boldsymbol{s}-\boldsymbol{h}_0}{\boldsymbol{s}-\boldsymbol{h}_0}\right].
\label{eq:likelihood}
\end{equation}
Here $\innerprod{\boldsymbol{s}-\boldsymbol{h}_0}{\boldsymbol{s}-\boldsymbol{h}_0}$ is the overlap between waveforms defined by the standard signal inner product \citep{Cutler1994}. This is the probability of the realisation of a noise signal $\boldsymbol{n}(t) = \boldsymbol{s}(t) - \boldsymbol{h}_0(t)$, assuming stationary, Gaussian noise.

We use uninformative priors on all the parameters: no existing constraints are used. We adopt uniform priors for all the parameters except: $M_\bullet$, $\zeta$ and $L_\infty$, which are positive definite, where we use a prior uniform in the logarithm of the parameter, and $\Theta\sub{K}$ and $\iota$, where we use a prior uniform in the cosine of the angle.

To assess the accuracy to which parameters can be determined, we must find the width of the posterior distribution. MCMC methods are widely used for inference problems; they are a class of algorithms used for integrating over complicated probability distributions. The parameter space is explored by constructing randomly a chain of samples, with an acceptance rate dependent upon their relative probabilities \citep{Metropolis1953,Hastings1970}. The distribution of points visited by the chain maps out the underlying distribution.

We employ the same semi-adaptive algorithm as was previously used in \citet{Berry2013}. In this scheme, there is an initial phase where the proposal distribution (used in the selection of new points) is adjusted to match the distribution of points previously accepted; this tailors the width of the proposal to the particular posterior under consideration \citep*{Haario1999}. The proposal is then fixed for the main phase to ensure ergodicity \citep{Roberts2007,Andrieu2008} and only accepted points from this phase are used to characterise the posterior. This allows us to efficiently recover posteriors for a wide range of bursts.

\section{Calculating event rates}\label{sec:Rates}

Having determined how to generate a waveform and extract the information from it, we must now consider how likely it is that such a waveform would be observed. We wish to calculate the event rate for EMRBs, the probability that there is an encounter between a CO, on an orbit described by eccentricity $e$ and periapse radius $r\sub{p}$, and the MBH. To do so we must build a model to describe the distribution of COs about the MBH. The number density of stars in the six-dimensional phase space of position and velocity is described by the distribution function (DF) $f$ \citep[section 4.1]{Binney2008}. We introduce approximate forms for the DF appropriate for describing the Galactic core in \secref{DF}. These are calibrated using the simulations of \citet{Alexander2009}, the parameters of which, together with the others used to describe the Galactic core, are given in \secref{GC-Param}. Having set the distribution of COs, we explain how to convert this to an event rate in \secref{e-rp}. In \eqnref{Gamma} we give an expression that relates the event rate for an orbit $\Gamma(e,\,r\sub{p})$ and the DF. There is one final consideration before we can calculate the total event rate: that there is an inner periapsis below which orbits become depopulated. This is carefully explained in \secref{inner-cut}. We consider tidal disruption and collisions, which we assume truncate the DF at a finite periapsis so that the event rate inside these cut-offs is zero. We also consider GW inspiral, which we assume alters the event rate by modifying the distribution of COs away from its relaxed state. With these inner cut-offs established, we have completely defined the event rate distribution. This can then give the probability of an EMRB and the total event rates, which are presented in \secref{no-events}.

\subsection{The distribution function}\label{sec:DF}

Following the work of \citet{Bahcall1976, Bahcall1977}, we assume that the DF within the Galactic core is only a function of the orbital energy \citep{Shapiro1978}. The energy per unit mass of the orbit is
\begin{equation}
\mathcal{E} = \dfrac{v^2}{2} - \dfrac{GM_\bullet}{r},
\end{equation}
where $v$ is the orbital velocity. The number of stars is
\begin{equation}
N = \int \dd^3r \int \dd^3v f(\mathcal{E}).
\end{equation}
Close to the centre of the Galactic core, the dynamics are dominated by the influence of the MBH as it is significantly more massive than the surrounding stars. Its radius of influence is
\begin{equation}
r\sub{c} = \dfrac{GM_\bullet}{\sigma^2},
\label{eq:r_c}
\end{equation}
where $\sigma^2$ is the line-of-sight velocity dispersion \citep{Frank1976}. We assume that the mass of stars enclosed within $r\sub{c}$ is greater than $M_\bullet$, which, in turn, is much greater than the mass of a typical star $M_\star$ \citep{Bahcall1976}. We define a reference number density $n_\star$ from the enclosed mass $m_\star(r)$ such that
\begin{equation}
m_\star(r\sub{c}) = \dfrac{4\pi r\sub{c}^3}{3}n_\star M_\star.
\end{equation}
Within the core, the DF can be calculated using the approximation of the Fokker--Planck formalism \citep[section 7.4]{Binney2008}. The population of bound stars is evolved numerically until a steady state is reached, whilst the unbound stars form a reservoir with an assumed Maxwellian distribution. Denoting a species of star by its mass $M$, the unbound DF is
\begin{equation}
f_M(\mathcal{E}) = \dfrac{C_M n_\star}{(2\pi\sigma_M^2)^{3/2}} \exp\left(-\dfrac{\mathcal{E}}{\sigma_M^2}\right),\quad\mathcal{E} > 0,
\label{eq:Unbound_DF}
\end{equation}
where $C_M$ is a normalisation constant.\footnote{$C_M$ determines the population ratios of species $M$ far from the MBH \citep{Alexander2009}.} If different stellar species are in equipartition, as assumed by \citet{Bahcall1976, Bahcall1977}, we expect
\begin{equation}
M \sigma_M^2 = M_\star \sigma_\star^2.
\end{equation}
However, if the unbound stellar population has reached equilibrium by violent relaxation, all mass groups are expected to have similar dispersions:
\begin{equation}
\sigma_M = \sigma_\star = \sigma,
\end{equation}
and we have equipartition of energy per unit mass \citep{Lynden-Bell1967}. This is assumed here following \citet{Alexander2009} and \citet{O'Leary2009}. The steady-state DF is largely insensitive to this choice \citep{Bahcall1977, Alexander2009}.

For bound orbits, the DF can be approximated as a power law \citep{Peebles1972}
\begin{equation}
f_M(\mathcal{E}) = \dfrac{k_M n_\star}{(2\pi\sigma^2)^{3/2}}\left(-\dfrac{\mathcal{E}}{\sigma^2}\right)^{p_M},\quad\mathcal{E} < 0.
\label{eq:Bound_DF}
\end{equation}
The exponent $p_M$ varies depending upon the mass of the object, determining mass segregation. For a system with a single mass component $p = 1/4$ \citep{Bahcall1976, Young1977}. The normalisation constant $k_M$ reflects the relative abundances of the different species.\footnote{For a single mass population ($p = 1/4$), $k = 2 C$ gives a fit correct to within a factor of two \citep{Bahcall1976,Keshet2009}; we assume that this holds for the dominant species of stars as, although it changes slightly with $p$, variation is small compared to errors introduced by fitting a simple power law \citep{Hopman2006, Alexander2009}.}

These cusp profiles should exist if the system has had sufficient time to become gravitationally relaxed. There is current debate about whether this may be the case, both for the GC and galaxies in general. This is discussed further in \apref{tauGC}. For concreteness, we assume that a cusp has formed. If a cusp has not formed, we expect there to be a shallower core profile, with fewer objects passing close to the MBH. Our results are therefore an upper bound on possible event rates \citep{Merritt2010a,Antonini2011,Gualandris2012}. 

\subsection{Model parameters}\label{sec:GC-Param}

We use the Fokker--Planck model of \citet{Hopman2006, Hopman2006a} and \citet{Alexander2009}. This includes four stellar species: MS stars, WDs, NSs and stellar mass BHs. Their properties are summarised in \tabref{HA}. The behaviour of the Fokker--Planck model has been verified by $N$-body simulations \citep{Baumgardt2004,Preto2010}.
\begin{table}
\begin{minipage}{\columnwidth}
 \centering
  \caption{Stellar model parameters for the Galactic core using the results of \citet{Alexander2009}. The MS star is used as a reference for the normalisation constants. The number fractions for unbound stars are estimates corresponding to a model of continuous star formation \citep{Alexander2005}; \citet{O'Leary2009} arrive at the same proportions.\label{tab:HA}}
  \begin{tabular}{@{} l D{.}{.}{2.1} D{.}{.}{1.3} D{.}{.}{1.1} D{.}{.}{1.3} @{}}
  \hline
   Star & \multicolumn{1}{c}{$M/M_\odot$} & \multicolumn{1}{c}{$C_M/C_\star$} & \multicolumn{1}{c}{$p_M$} & \multicolumn{1}{c}{$k_M/k_\star$\footnote{\citet*{Toonen2009}}} \\
 \hline
 MS & 1.0 & 1 & -0.1 & 1 \\
 WD & 0.6 & 0.1 & -0.1 & 0.09 \\
 NS & 1.4 & 0.01 & 0.0 & 0.01  \\
 BH & 10 & 0.001 & 0.5 & 0.008 \\
\hline
\end{tabular}
\end{minipage}
\end{table}
The steeper power law for BHs means they segregate about the MBH.\footnote{Extrapolating, they would dominate in place of MS stars for radii $r < 10^{-4}r\sub{c}$.}

Binaries may form in the Galactic core, encouraged by its high stellar density \citep{O'Leary2009}. However, the binary fraction is still expected to be small \citep{Hopman2009}. Binaries are also disrupted by the MBH for periapses smaller than
\begin{equation}
r\sub{B} \simeq \left(\dfrac{M_\bullet}{M_1 + M_2}\right)^{1/3}a\sub{B},
\end{equation}
where $M_1$ and $M_2$ are the masses of the binary's components, and $a\sub{B}$ is the binary's semi-major axis, cf.\ \eqnref{Tidal} below. Thus, we ignore the possible presence of binaries.

We assume that $M_\bullet = (4.31 \pm 0.36) \times 10^6 M_\odot$ \citep{Gillessen2009} and $\sigma = (103 \pm 20)\units{km\,s^{-1}}$ \citep{Tremaine2002}. This gives a core radius of $r\sub{c} = (1.7 \pm 0.7)\units{pc}$. Using the results of \citet{Ghez2008} we would expect the total mass of stars in the core to be $m_\star(r\sub{c}) = 6.4 \times 10^6 M_\odot$, which is within $5\%$ of the value obtained similarly from \citet{Genzel2003}. This gives a reference stellar density of $n_\star = 2.8 \times 10^5\units{pc^{-3}}$.

\subsection{The event rate in terms of eccentricity and periapsis}\label{sec:e-rp}

We characterise orbits by their eccentricity $e$ and periapse radius $r\sub{p}$. The latter, unlike the semimajor axis, is always well defined regardless of eccentricity. For Keplerian orbits, the energy $\mathcal{E}$ and angular momentum $\mathcal{J}$ per unit mass are entirely characterised by these parameters
\begin{equation}
\label{eq:Energy_ecc}
\begin{split}
\mathcal{E} & = -\dfrac{GM_\bullet(1 - e)}{2r\sub{p}}; \\
\mathcal{J}^2 & = GM_\bullet(1 + e)r\sub{p}.
\end{split}
\end{equation}
The DF is defined per element of phase space: it is necessary to change variables from position and velocity to eccentricity and periapsis. We decompose the velocity into three orthogonal components: radial $v_r$, azimuthal $v_\phi$ and polar $v_\theta$. We assume that the core is spherically symmetric \citep{Genzel2003, Schodel2007}, therefore we are interested in the combination
\begin{equation}
v_\perp^2 = v_\phi^2 + v_\theta^2 = v^2 - v_r^2.
\end{equation}
Under this change of variables
\begin{equation}
\dd^3v = \dd v_r \dd v_\phi \dd v_\theta \rightarrow 2\pi v_\perp \,\dd v_r \,\dd v_\perp.
\end{equation}
The specific energy and angular momentum are given by
\begin{equation}
\begin{split}
\mathcal{E} & = \dfrac{v_r^2 + v_\perp^2}{2} - \dfrac{GM_\bullet}{r}; \\
\mathcal{J}^2 & = r^2 v_\perp^2.
\end{split}
\end{equation}
Combining these with our earlier expressions in terms of $e$ and $r\sub{p}$,
\begin{align}
v_\perp^2 = {} & \dfrac{GM_\bullet(1 + e)r\sub{p}}{r^2}, \\*
v_r^2 = {} & GM_\bullet\left[\dfrac{2}{r} - \dfrac{(1 - e)}{r\sub{p}} - \dfrac{(1 + e)r\sub{p}}{r^2}\right].
\end{align}
From the latter we can verify that the turning points of an orbit occur at
\begin{equation}
r = r\sub{p}, \: \dfrac{1+e}{1-e}r\sub{p};
\end{equation}
the periapse is the only turning point for orbits with $e > 1$. Since we now have expressions for $\{v_r, v_\perp\}$ in terms of $\{e, r\sub{p}\}$, we can 
rewrite our velocity element as
\begin{equation}
\dd^3v \rightarrow \dfrac{\pi e}{v_rr\sub{p}}\left(\dfrac{GM_\bullet}{r}\right)^2\,\dd e \,\dd r\sub{p}.
\end{equation}
As a consequence of our assumed spherical symmetry, 
the phase space volume element can be expressed as
\begin{equation}
\dd^3r\dd^3v \rightarrow \dfrac{4\pi^2(GM_\bullet)^2e}{v_rr\sub{p}}\,\dd r\,\dd e \,\dd r\sub{p}.
\end{equation}
The number of stars in an element $\dd r\,\dd e\,\dd r\sub{p}$ is
\begin{equation}
n(r, e, r\sub{p}) = \dfrac{4\pi^2(GM_\bullet)^2e}{v_rr\sub{p}}f(\mathcal{E}).
\end{equation}

From this, we can construct the expected number of stars on orbits defined by $\{e, r\sub{p}\}$. The number of stars found in a small radius range $\delta r$ with given orbital properties is
\begin{equation}
n(r, e, r\sub{p})\delta r = N(e, r\sub{p}; r)\dfrac{\delta t}{P(e, r\sub{p})},
\end{equation}
where $N(e, r\sub{p}; r)$ is the total number of stars with orbits given by $\{e, r\sub{p}\}$ defined at $r$, $\delta t$ is the time spent in $\delta r$ and $P(e, r\sub{p})$ is the period of the orbit. We defer the definition of this time for unbound orbits for now. The time spent in the radius range is
\begin{equation}
\delta t = 2\dfrac{\delta r}{v_r},
\end{equation}
where the factor of $2$ accounts for inward and outward motion. Hence,
\begin{align}
N(e, r\sub{p}; r) = {} & \recip{2} v_r P(e, r\sub{p}) n(r, e, r\sub{p}) \nonumber \\
 = {} & \dfrac{2\pi^2(GM_\bullet)^2 e P(e, r\sub{p})}{r\sub{p}}f(\mathcal{E}).
\end{align}
The right hand side is independent of position, subject to the constraint that the radius is in the allowed range for the orbit $r\sub{p} \leq r \leq (1+e)r\sub{p}/(1-e)$, and so $N(e, r\sub{p}) \equiv N(e, r\sub{p}; r)$. This is a consequence of the DF being dependent only upon a constant of the motion.\footnote{See \citet[equation 9]{Bahcall1976} for a similar result.}

If a burst of radiation is emitted each time a star passes through periapse, the event rate for burst emission from orbits with parameters $\{e, r\sub{p}\}$ is given by
\begin{align}
\Gamma(e, r\sub{p}) = {} & \dfrac{N(e, r\sub{p})}{P(e, r\sub{p})} = \dfrac{2\pi^2(GM_\bullet)^2 e}{r\sub{p}}f(\mathcal{E}).
\label{eq:Gamma}
\end{align}
The orbital period drops out from the calculation, so we do not have to worry about an appropriate definition for unbound orbits.



To generate a representative sample for the orbital parameters $e$ and $r\sub{p}$, we use $\Gamma(e, r\sub{p})\dd e\, \dd r\sub{p}$ as the rate for a Poisson distribution.

The total event rate can be found by integrating $\Gamma(e,\,r\sub{p})$, but before we can do this we must set the limits of the integral. The maximum periapse is set by the limit of detectability. It is particular to the detector. The inner periapse is set by physical processes. These are explained in the following section.

\subsection{The inner cut-off}\label{sec:inner-cut}

From \eqnref{Gamma} we see that the event rate is highly sensitive to the smallest value of the periapsis. Ultimately, the orbits cannot encroach closer to the MBH than its last stable orbit. This depends upon the spin of the MBH, but is of the order of its Schwarzschild radius. Before we reach this point, there are other processes that may intervene to deplete the orbiting stars. Our treatment of these is approximate, but should produce reasonable estimates. We consider three processes: tidal disruption by the MBH (\secref{Tidal}), GW inspiral (\secref{GW-in}) and collisional disruption (\secref{Collision}). Tidal disruption imposes a definite (albeit approximate) cut-off, while the others use statistical arguments. For these methods, we need to define a reference time-scale for relaxation. This is done in \secref{Relax}, with further details found in \apref{time-scale}.

\begin{figure*}
\begin{center}
   \subfigure[{Main sequence stars}]{\includegraphics[width=0.4\textwidth]{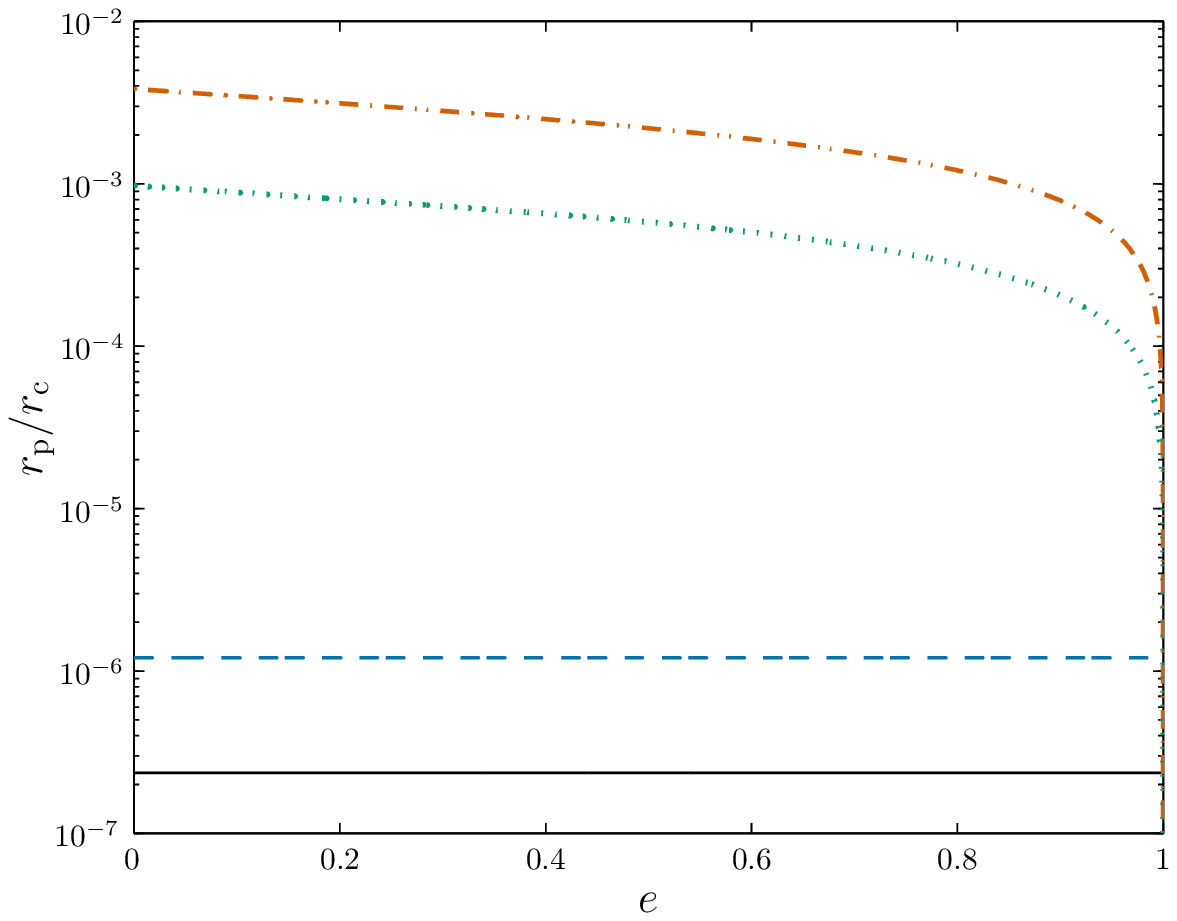}} \quad 
   \subfigure[{White dwarfs}]{\includegraphics[width=0.4\textwidth]{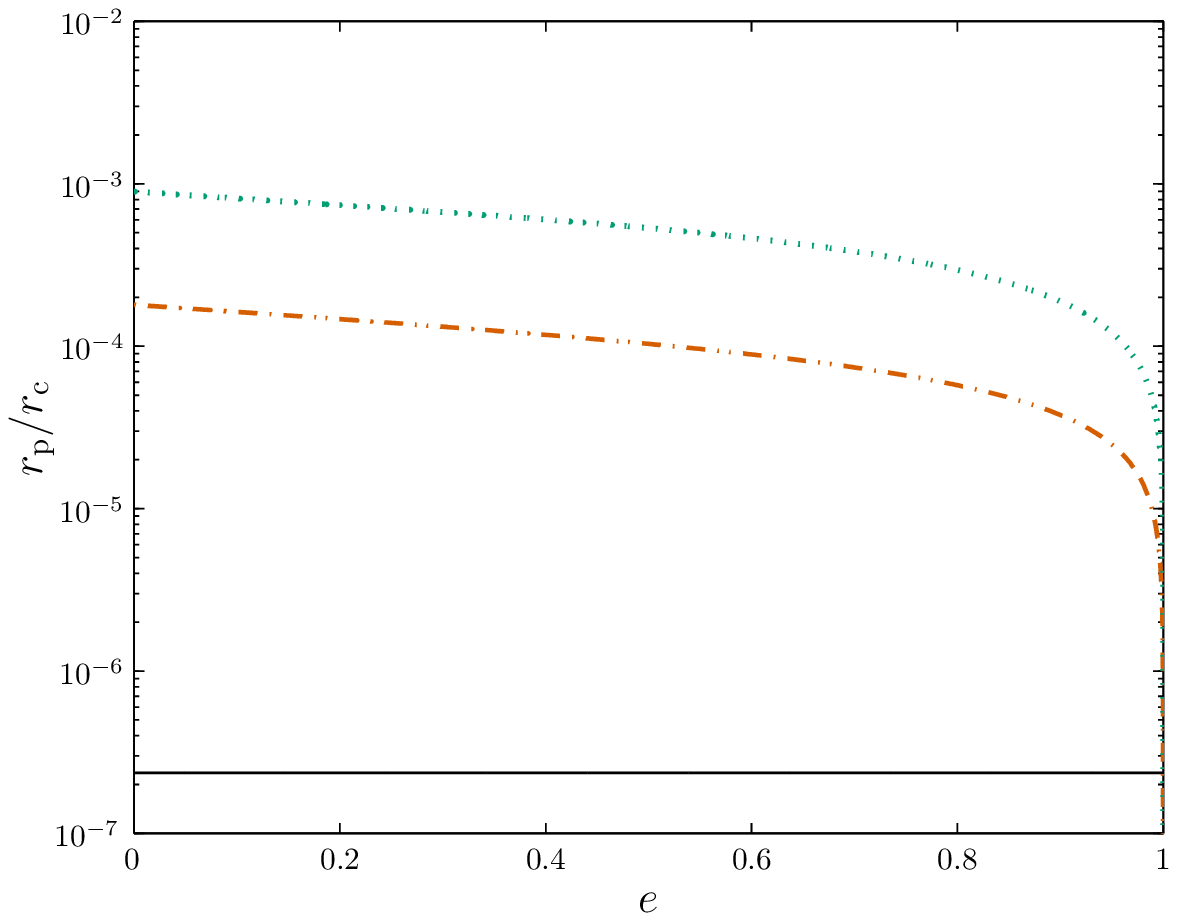}} \\
   \subfigure[{Neutron stars}]{\includegraphics[width=0.4\textwidth]{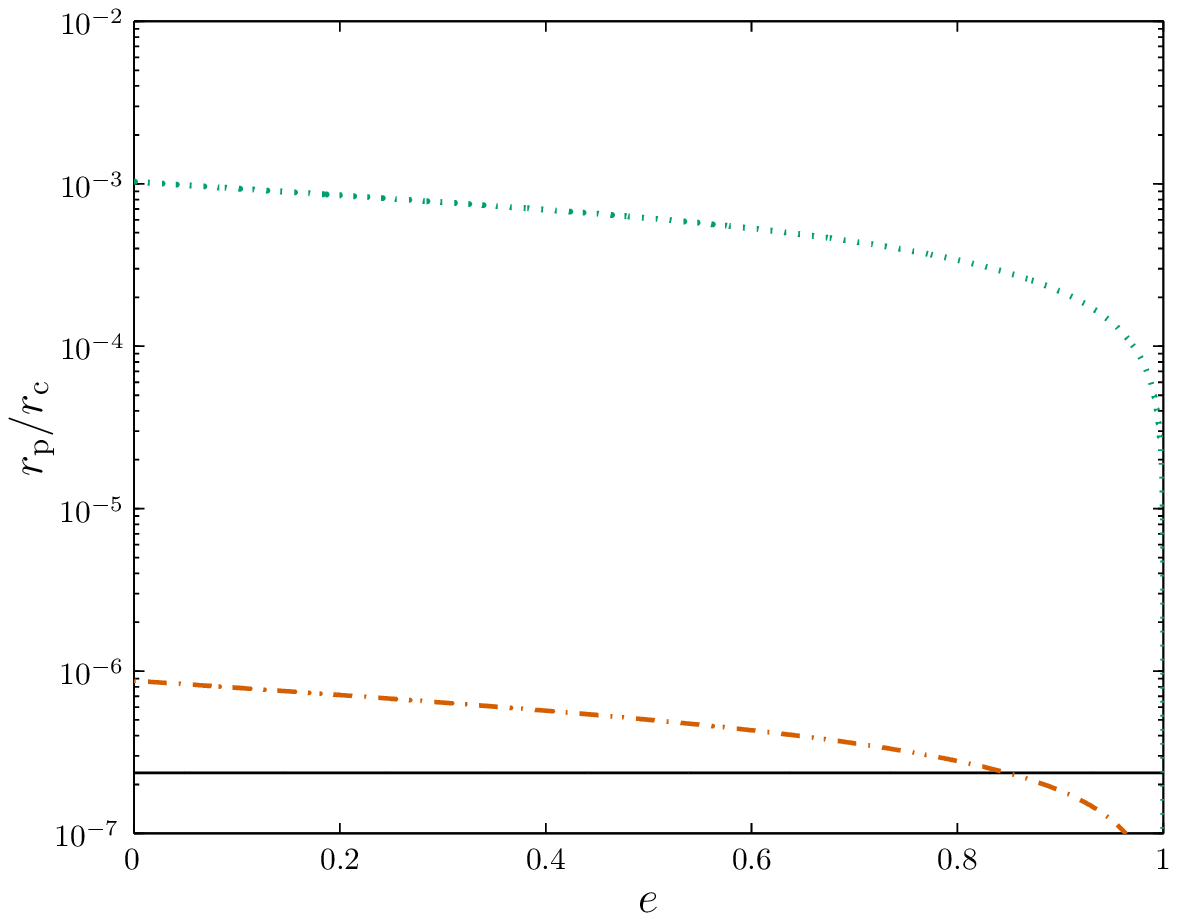}} \quad
   \subfigure[{Black holes}]{\includegraphics[width=0.4\textwidth]{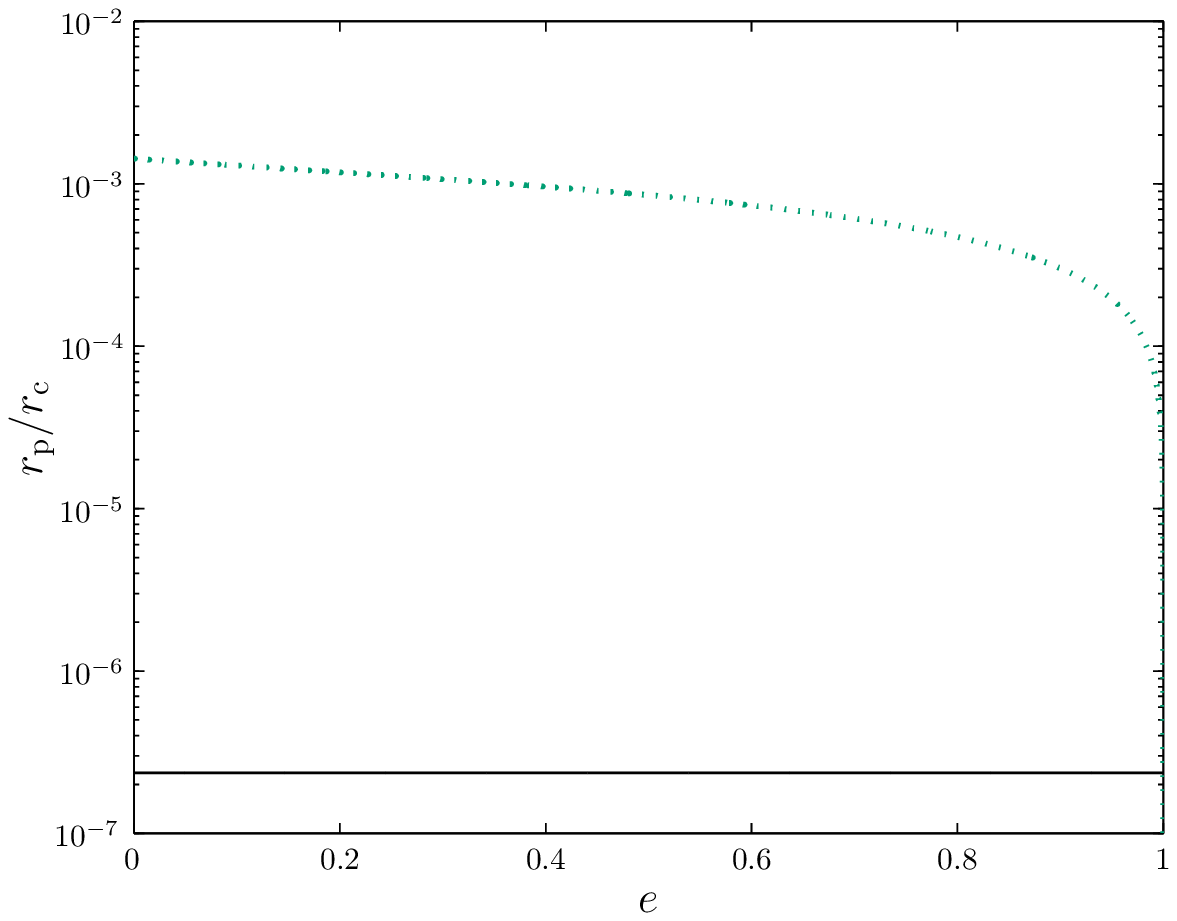}}
\caption{Inner cut-off radii for the GC as a function of eccentricity. The solid line shows the Schwarzschild radius of the MBH; this gives an indication of the innermost possible orbit which actually varies with MBH spin as well as orbital eccentricity and inclination. The dashed line shows the tidal radius which is a hard cut-off inside of which there should be no undisrupted stars. The dot--dashed line shows the collisional cut-off which is a statistical cut-off inside of which we do not expect any stars. The dotted line shows the transition to the GW-dominated inspiral regime; inside of this we expect inspiralling stars in place of the relaxed distribution.\label{fig:Cuts}}
  \end{center}
\end{figure*}

The calculated inner cut-offs for the four stellar species across the range of bound orbits are shown in \figref{Cuts}.
The tidal and collisional disruption cut-offs are hard boundaries, inside of which we assume that there are no bursting sources. The transition to the GW inspiral dominated regime marks the end of the relaxed distribution of stars; inside of this there are only inspiralling stars.

\subsubsection{Tidal disruption}\label{sec:Tidal}

Tidal forces from the MBH can disrupt stars. This occurs at the tidal radius
\begin{equation}
r\sub{T} \simeq \left(\dfrac{M_\bullet}{M}\right)^{1/3}R_M,
\label{eq:Tidal}
\end{equation}
where $R_M$ is the radius of the star \citep{Hills1975, Rees1988, Kobayashi2004}.\footnote{See \citet{Kesden2012} for a general relativistic treatment.} Any star on an orbit with $r\sub{p} < r\sub{T}$ is disrupted in the course of its orbit. Parametrizing orbits by their periapsis allows us to easily determine which stars should be disrupted. We do not include the full effects of the loss cone \citep{Frank1976, Lightman1977, Cohn1978} as these were not incorporated into the Fokker--Planck calculations \citep{Hopman2009}.\footnote{The loss cone is a region in velocity space where orbits are depleted because stars are disrupted more rapidly than they can be replenished by two-body scattering.} The effect of the loss cone should be small, only modifying the DF by a logarithmic term \citep{Lightman1977, Bahcall1977, Cohn1978}. Its effects are diluted by resonant relaxation \citep{Hopman2007,Toonen2009,Merritt2011}. Furthermore, the loss cone could be refilled by the wandering within the core of the MBH because of perturbations from the inhomogeneities in the stellar potential \citep{Sigurdsson1997,Chatterjee2002,Merritt2007}.

Tidal disruption is significant for MS stars since they are least dense: calculated in this way, only MS stars are tidally disrupted outside of the MBH's event horizon \citep{Sigurdsson1997}. The tidal radius defines the cut-off for periapsis of high-eccentricity ($e \ga 1$) orbits \citep{Lightman1977}.

\subsubsection{Relaxation time-scale}\label{sec:Relax}

The motion of a star is determined not only by the dominant influence of the central MBH, but also by the other stars. The gravitational potential of the stars may be split into two components: a smooth background representing the average distribution of stars, and statistical fluctuations from random deviations in the stellar distribution because of individual stellar motions. The former only contributes to the stars' orbits: we neglect this since we are more interested in the influence of the MBH. The latter may be approximated as a series of two-body encounters. These lead to scattering, in a manner much like Brownian motion \citep{Bekenstein1992,Maoz1993,Nelson1999}.

The two-body interactions mostly lead to small deflections. Over time, these may accumulate into a significant change in the dynamics. The relaxation time-scale characterises the time taken for this to happen \citep[section 1.2.1]{Binney2008}. It therefore quantifies the time over which an orbit may be repopulated by scattering. There are a variety of definitions for the relaxation time-scale. For a system with a purely Maxwellian distribution, the time-scale has the form
\begin{equation}
\tau\sub{R}\super{Max} \simeq \kappa\dfrac{\sigma^3}{G^2M_\star^2 n_\star\ln\Lambda},
\label{eq:tauMaxwell}
\end{equation}
where the Coulomb logarithm is $\ln\Lambda = \ln(M_\bullet/M_\star)$ \citep{Bahcall1976}, and $\kappa$ is a dimensionless number. In his pioneering work, \citet{Chandrasekhar1941, Chandrasekhar1960} defined the time-scale as the period over which the squared change in energy was equal to the kinetic energy squared; this gives $\kappa = 9/16\sqrt{\pi} \simeq 0.32$. Subsequently, \citet{Chandrasekhar1941a} described relaxation statistically, treating fluctuations in the gravitational field probabilistically; this gives $\kappa = 9/2(2\pi)^{3/2} \simeq 0.29$. \citet{Bahcall1977} define a reference time-scale from their Boltzmann equation with $\kappa = 3/4\sqrt{8\pi} \simeq 0.15$; this is equal to the reference time-scale defined as the reciprocal of the coefficient of dynamical friction by \citet{Chandrasekhar1943a, Chandrasekhar1943}. \citet{Spitzer1958} define a reference time-scale from the gravitational Boltzmann equation of \citet{Spitzer1951} where $\kappa = \sqrt{2}/\pi \simeq 0.45$. Following \citet{Spitzer1971}, \citet[section 7.4.5]{Binney2008} estimate the time-scale from the velocity diffusion coefficient of the Fokker--Planck equation yielding $\kappa \simeq 0.34$.

All these approaches yield consistent values, suggesting, as a first approximation, any is valid. We follow the classic treatment of \citet[chapter 2]{Chandrasekhar1960} which is transparent in its assumptions, adapting from a Maxwellian distribution of velocities to one derived from the DFs \eqnref{Unbound_DF} and \eqnref{Bound_DF}; this makes the model self-consistent. The derivation of the relaxation time-scale along with a discussion of its shortcomings is included in \apref{time-scale}. An average time-scale for the entire system $\overline{\tau\sub{R}}$ is defined in \eqnref{system-relax}, and an average for an orbit $\left\langle\tau\sub{R}\right\rangle$ is defined in \eqnref{orbital-relax}. 

Since there is uncertainty in the astrophysical parameters, we will not be concerned by small discrepancies in the numerical prefactor that result from the simplifying approximations of this approach. We defer any investigation of the consequences of using an alternative formulation for future work, as differences may well be negligible whilst the computation is complicated.

Two-body interactions lead to diffusion in both energy and angular momentum. When considering a single (bound) orbit, over a relaxation time-scale the energy changes by order of itself while the angular momentum changes by the angular momentum of a circular orbit with that energy $\mathcal{J}\sub{circ}(\mathcal{E})$ \citep{Lightman1977, Rauch1996, Hopman2005, Madigan2011}:\footnote{$\mathcal{J}\sub{circ}(\mathcal{E})$ is the maximum value for orbits of that energy.}
\begin{equation}
\left(\dfrac{\Delta\mathcal{E}}{\mathcal{E}}\right)^{2} \approx \left[\dfrac{\Delta \mathcal{J}}{\mathcal{J}\sub{circ}(\mathcal{E})}\right]^{2} \approx \dfrac{t}{\tau\sub{R}}.
\label{eq:diffuse-relax}
\end{equation}
We may define another angular momentum relaxation time-scale as the time taken for the angular momentum to change by order of itself \citep{Merritt2011}
\begin{align}
\tau_\mathcal{J} = {} & \left[\dfrac{\mathcal{J}}{\mathcal{J}\sub{circ}(\mathcal{E})}\right]^2\tau\sub{R} = \left(1 - e^2\right) \tau\sub{R}.
\label{eq:J-time}
\end{align}
This can be much shorter than the energy relaxation time-scale: diffusion in angular momentum can proceed more rapidly than diffusion in energy.

\subsubsection{Gravitational wave inspiral}\label{sec:GW-in}

Stars orbiting the MBH continually emit gravitational radiation; this carries away energy and angular momentum, causing the stars to inspiral. Using the analysis of \citet{Peters1963} and \citet{Peters1964} for Keplerian binaries, it is possible to define a characteristic inspiral time-scale from the rate of change of energy. For consistency with the relaxation time-scale, we define this as \citep{MiraldaEscude2000, Merritt2011}
\begin{equation}
\tau\sub{GW} \simeq \mathcal{E}\left\langle\diff{\mathcal{E}}{t}\right\rangle^{-1},
\label{eq:tGW-def}
\end{equation}
where the term in angular brackets is the orbit-averaged rate of energy radiation. Using \eqnref{Energy_ecc} and equation 16 of \citet{Peters1963},
\begin{align}
\tau\sub{GW} \simeq {} & \dfrac{5}{64}\dfrac{c^5r\sub{p}^4}{G^3MM_\bullet\left(M + M_\bullet\right)}\dfrac{(1+e)^{7/2}}{(1-e)^{1/2}} \nonumber \\*
 {} & \times {} \left(1+\dfrac{73}{24}e^2 + \dfrac{37}{96}e^4\right)^{-1} \\
 \approx {} & \dfrac{5}{64}\dfrac{c^5r\sub{p}^4}{G^3MM_\bullet^2}\dfrac{(1+e)^{7/2}}{(1-e)^{1/2}}\left(1+\dfrac{73}{24}e^2 + \dfrac{37}{96}e^4\right)^{-1}.
\end{align}
For comparison, the total time taken for the inspiral, if undisturbed, is given in \eqnref{Bound_inspiral}.

The time-scale associated with changes in angular momentum is \citep{Peters1964}
\begin{align}
\tau_{\mathrm{GW},\, \mathcal{J}} \simeq {} & \mathcal{J}\left\langle\diff{\mathcal{J}}{t}\right\rangle^{-1} \\
 \simeq {} & \dfrac{5}{32}\dfrac{c^5r\sub{p}^4}{G^3MM_\bullet\left(M + M_\bullet\right)}\dfrac{(1+e)^{5/2}}{(1-e)^{3/2}}\left(1+\dfrac{7}{8}e^2\right)^{-1} \\
 \approx {} & \dfrac{5}{32}\dfrac{c^5r\sub{p}^4}{G^3MM_\bullet^2}\dfrac{(1+e)^{5/2}}{(1-e)^{3/2}}\left(1+\dfrac{7}{8}e^2\right)^{-1}.
\end{align}
This is always greater than the energy time-scale; hence, we only consider changes in energy from GW emission as important for evolution of the system \citep{Hopman2005}.

Unbound stars only undergo a single periapse passage and only radiate one burst of radiation; we therefore neglect any evolution in their orbital parameters.\footnote{Changes are only important for very high eccentricity orbits (\apref{Unbound}). These are high energy and are exponentially suppressed because of the Boltzmann factor in \eqnref{Unbound_DF}.}

The $(1-e)^{-1/2}$ dependence of $\tau\sub{GW}$ for bound orbits connects the two regimes. The rate of change of energy goes to zero as a consequence of assuming the orbital parameters do not change over the course of an orbit. It is a valid approximation since the large mass-ratio ensures a slow evolution of the system (\apref{Unbound}).

When comparing with the relaxation time-scale we are comparing rates of change, with the shorter time-scale highlighting the more rapid process that dominates the evolution \citep{Amaro-Seoane2007}. We therefore compare $\tau\sub{GW}$ with the orbital relaxation time-scale $\tau_\mathcal{J}$ \citep{Merritt2011}. Orbits with $\tau\sub{GW} < \tau_\mathcal{J}$ become depleted by GW emission faster than they are replenished by scattering. The cusp does not extend to these orbits. Yet, these orbits are not totally depopulated as an object may pass through during its inspiral from greater periapse and eccentricity. In their calculations, \citet{Hopman2007} did not include these inspiralling stars as potential burst sources. We calculate the density of stars in this region by following the evolution of inspirals beginning at the inner edge of the cusp (where the two time-scales are equal), weighting by the rate of change of the periapse and eccentricity in each element of $e$--$r\sub{p}$ space \citep{Peters1964}. The net effects are the high-eccentricity distributions of MS stars, WDs and NSs are relatively unchanged from their cusp states, but the BH distribution is significantly depleted.

\subsubsection{Collisions}\label{sec:Collision}

As a consequence of the high densities in the Galactic core, stars may undergo a large number of close encounters with other stars \citep{Cohn1978}. These may lead to their destruction. MS stars, WDs and NSs may be pulled apart by tidal forces if they stray too close to a more massive object. As MS stars are diffuse, they would not tidally disrupt another star \citep{Murphy1991,Freitag2005}. Close encounters would result in some mass transfer; the cumulative effect of $20$--$30$ grazing collisions could destroy an MS star \citep{Freitag2006}. The number of collisions a star undergoes in a time interval $\delta t$ is
\begin{equation}
\delta K = n(r) A v(r,e,r\sub{p})\delta t,
\end{equation}
where $A$ is the collisional cross-sectional area. For tidal disruption, where the encounter is with a collapsed object (WD, NS or BH), we set $A = \pi r_{\mathrm{T},\,{M'}}^2$, where $r_{\mathrm{T},\,{M'}}$ is the appropriate tidal radius: like \eqnref{Tidal} but with $M_\bullet$ replaced with the mass of the collapsed object $M'$. For collisions between MS stars, the cross-sectional area is simply the geometric $A = \pi R_\star^2$.\footnote{Here we assume that the relative velocity of the colliding stars is much greater than the escape velocity of the star so we may neglect the effects of gravitational focusing.}

For circular orbits, we can find the radius at which collisions lead to disruptions by setting $\delta K = 1$ for tidal disruption or $\delta K = 20$ for grazing collisions, and $\delta t = \overline{\tau_{\mathrm{R},\,M}}$. We use the system average relaxation time-scale for the species of mass $M$ as this is the time over which stars are replenished from the reservoir. For non-circular orbits, we must consider variation with position. Using $\delta r = v_r \delta t$, and then converting to an integral, for bound orbits
\begin{equation}
K = 2 A \dfrac{\overline{\tau_{\mathrm{R},\,M}}}{P(r\sub{p},e)}\intd{r\sub{p}}{(1+e)r\sub{p}/(1-e)}{n(r)\dfrac{v(r,e,r\sub{p})}{v_r(r,e,r\sub{p})}}{r},
\label{eq:collision-K}
\end{equation}
where $P$ is the period of the orbit. Again we set $K = 1$ or $K = 20$, and then numerically solve \eqnref{collision-K} to find the orbits for which stars are disrupted within $\overline{\tau_{\mathrm{R},\,M}}$. For unbound orbits we are only interested in stars that would become disrupted before their periapse passage, so
\begin{equation}
K = A \intd{r\sub{p}}{r\sub{c}}{n(r)\dfrac{v(r,e,r\sub{p})}{v_r(r,e,r\sub{p})}}{r},
\end{equation}
assuming that the stars in the reservoir external to the core are unlikely to undergo close collisions.

Orbits within the collisional cut-off are assumed to be depopulated and do not contribute to the event rate. Our treatment is similar to that of \citet{Hopman2007}, but they only considered collisions between MS stars. Collisions provide the cut-off for bound MS stars and are significant for bound WDs.

\section{Results}\label{sec:Gal-Results}

\subsection{Number of events}\label{sec:no-events}

As a first approximation for the number of events expected in a $2\units{yr}$ mission lifetime, we numerically integrated the event rate. This estimate is denoted by $\mathcal{N}_2\super{int}$. The lower limit on $r\sub{p}$ was set to be the largest of the tidal cut-off, the collisional cut-off or the MBH's Schwarzschild radius.\footnote{The transition to the GW inspiral regime is not a cut-off, but reflects a change in the form of the stellar distribution; hence, it is not included amongst the other inner periapses as a lower limit.} The Schwarzschild radius $r\sub{S} = 2 r\sub{g}$ is used as a proxy for an averaged innermost orbit's periapse; the innermost parabolic orbit for non-spinning MBHs has $r\sub{p} = 4r\sub{g}$ and the innermost parabolic orbit for a maximally spinning MBH has $r\sub{p} = r\sub{g}$ for a prograde equatorial orbit and $r\sub{p} = (3 + 2\sqrt{2})r\sub{g}$ for a retrograde equatorial orbit. The upper limit was the detection threshold as determined from \eqnref{SNR-power-law}. The lower bound on eccentricity was set to $0.9$, below which we do not trust the parabolic approximation for burst waveforms; since the DF decays exponentially with eccentricity for unbound orbits, the upper limit does not influence our results.

To obtain a more accurate estimate, we performed $2 \times 10^4$ mission realisations. For each mission, we randomly selected a set of parameters to describe the MBH, and then picked orbits with probabilities defined by their event rates. The orbital position of the \textit{LISA} detector was also chosen randomly.\footnote{We use the same condition on the initial orientation as in \citet{Berry2013}, following \citet{Cutler1998}. This does not qualitatively influence our results.} The SNRs of the resulting bursts were calculated and a detection was recorded if $\rho > 10$. By averaging the number of events per mission, we can estimate the expected number of bursts we would detect. This is denoted by $\mathcal{N}_2\super{run}$.

\begin{figure}
\begin{center}
   \includegraphics[width=0.45\textwidth]{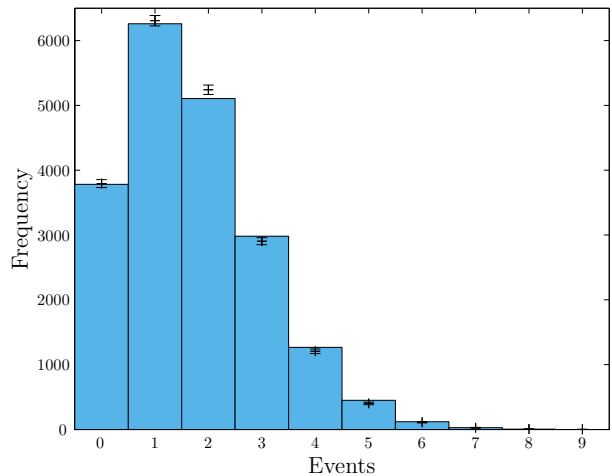}
\caption{Calculated number of detectable EMRBs over a two-year mission. The histogram shows the number of events for $2 \times 10^4$ realisations. The points show a Poisson distribution with a mean set by $\mathcal{N}_2\super{int}$.}
\label{fig:Event-no}
\end{center}
\end{figure}

The calculated numbers of events are shown in \tabref{Rates}.
\begin{table}
\begin{minipage}{\columnwidth}
 \centering
  \caption{Expected number of events per two-year mission. $\mathcal{N}_2\super{int}$ is an estimate using the average SNR--periapsis scaling and $\mathcal{N}_2\super{run}$ is calculated by averaging results from $2 \times 10^4$ mission realisations.\label{tab:Rates}}
  \begin{tabular}{@{} l D{,}{\times}{3.4} D{,}{\times}{3.4} @{}}
  \hline
   Star & \multicolumn{1}{c}{$\mathcal{N}_2\super{int}$} & \multicolumn{1}{c}{$\mathcal{N}_2\super{run}$} \\
 \hline
 MS & 9.5,10^{-4} & 1.3,10^{-3} \\
 WD & 1.0,10^{-2} & 1.0,10^{-2} \\
 NS & 5.0,10^{-1} & 5.0,10^{-1}  \\
 BH & 1.2,10^{0} & 1.2,10^{0} \\
\hline
Total & 1.7,10^{0} & 1.7,10^{0} \\
\hline
\end{tabular}
\end{minipage}
\end{table}
The two approaches are in good agreement indicating that the average relation \eqnref{SNR-power-law} is sufficiently accurate for this type of calculation, and that the Schwarzschild radius is a reasonable absolute inner cut-off averaged over all MBH spins. The total number of events per mission is plotted in \figref{Event-no}.
This is consistent with being Poisson distributed as expected.

Only BHs and NSs contribute to the event rate significantly. Only MS stars have a non-negligible (relative) contribution from unbound orbits. The event rates are not high, but there is an $\sim 4/5$ ($81\%$) chance of observing at least one burst in a mission.

The overall rates are similar to those presented in \citet{Hopman2007}. The MS rate is lower because of a larger collisional cut-off. This also influences the WD rate, but the overall rate is little changed. The NS rate is enhanced because of the inclusion of bursts from inspiralling objects. The physics for BHs is least changed; the (small) difference in the event rate is partly a consequence of our more realistic SNRs.

\subsection{Information content}

We wish to quantify what we could learn over a mission about the MBH's mass and spin. We use the parameter set $\boldsymbol{\lambda}_\bullet = \{\ln (M_\bullet/M_\odot), a_\ast, \cos \Theta\sub{K}, \Phi\sub{K}\}$ as each of these has a uniform prior.

The information carried by a burst is encoded in its posterior probability distribution. This can be recovered using an MCMC as explained in \secref{param-est-MCMC}. We ran MCMCs for bursts from the first $100$ of our mission realisations that had periapses $r\sub{p} < 16 r\sub{g}$. There were a total of $96$ interesting bursts ($57$ from BHs and $39$ from NSs) across $63$ missions. For examples of recovered posterior distributions, see \citet{Berry2013,Berry2013a}.\footnote{The former shows posteriors for Galactic bursts where the CO is $10 M_\odot$. The latter shows posteriors for extragalactic bursts and a $10 M_\odot$ CO. The CO mass is degenerate with the distance to the source; hence, the extragalactic bursts give an indication of what happens when the CO mass is reduced (although changing from a $10 M_\odot$ BH to a $1.4 M_\odot$ NS is not as extreme as moving to another galaxy).} Ideally, we would use information from all detectable bursts, but this would be computationally expensive and we do not expect to glean much useful information from orbits with larger periapses.

During an individual mission there may be either zero, one or multiple bursts of interest. In the first case, we learn nothing. In the second, we have only to consider the posterior from our MCMC. In the third, we must combine the posteriors of all the bursts. This is easy in theory: as the priors are uniform we have only to multiply the individual posteriors,
\begin{equation}
p(\boldsymbol{\lambda}_\bullet|\{\boldsymbol{s}_i(t)\}\sub{mission}) = \prod_i p(\boldsymbol{\lambda}_\bullet|\boldsymbol{s}_i(t)),
\end{equation}
where $\{\boldsymbol{s}_i(t)\}\sub{mission}$ is the set of bursts for the mission. However, since we have a sampled posterior rather than an analytic function, this is difficult in practice.

The simplest thing to do is bin the points and then multiply the numbers in each bin together (dividing by the area of the bin to convert back to a probability density). The question is then what is an appropriate bin size? Bins that are too big give insufficient resolution, whilst those that are too small may not encompass any sampled points.

One means of creating bins with sizes that reflect the structure of the distribution is using a $k$-d tree. This is a type of binary space partitioning tree \citep[sections 5.2, 12.1 and 12.3]{Berg2008}; it is constructed finding the median in one dimension and then splitting the parameter space into two at this location. This process is then repeated in another dimension, then another, and so on recursively until the desired number of partitions, known as leaves, have been created. Constructing a $k$-d tree from a sampled probability distribution creates smaller leaves in the regions of high probability, which are of most interest, and larger leaves in the areas of low probability \citep{Weinberg2012}.

Taking each burst posterior in turn, we construct a $k$-d tree using the two-step method \citep*{Sidery2013,Berry2013a}: the data are (randomly) divided in two parts --- the first half is used to create the leaves and the second half is used to populate the leaves. This reduces biasing due to fluctuations in the sampled data. We use this tree to bin the other posteriors and multiply the totals together. This gives us one estimate for the combined posterior for each of the input bursts. We resample the final distributions (sampling each leaf uniformly) to create sets of points that can be treated in the same way as the output from an MCMC.

The precision to which a parameter can be constrained may be quantified by the width of the distribution. We shall use the standard deviation $\sigma\sub{SD}$ and the half-width of the $68$-percentile range constructed from one-dimensional $k$-d trees $\sigma_{0.68}$. This is constructed by adding together the smallest leaves of the tree until they encompass a total probability of $0.68$. The widths coincide for Gaussian posteriors.

\begin{figure}
\begin{center}
   \includegraphics[width=0.45\textwidth]{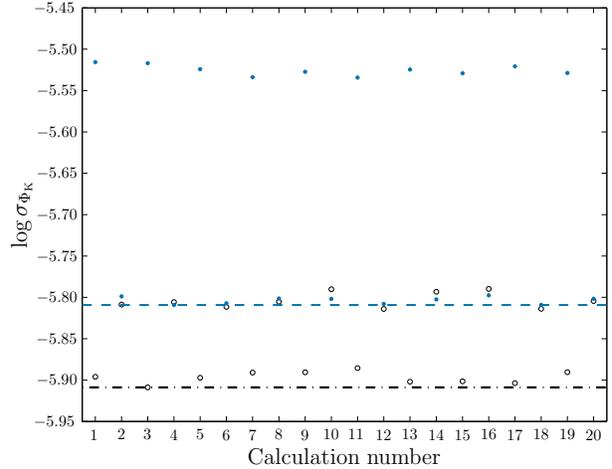}
\caption{Example of calculated values for the standard deviation $\sigma\sub{SD}$ (indicated by the solid points) and the $68$-percentile half width $\sigma\sub{0.68}$ (indicated by the open points) that show clear biasing. The odd numbered calculations use the $k$-d tree constructed from the first burst for binning and the even numbered calculations use the tree from the second. Both of these processes were repeated $10$ times to check there were no numerical problems. The values used as final results for $\sigma\sub{SD}$ and $\sigma\sub{0.68}$ are indicated by the dashed and dot--dashed lines respectively. The posterior used here is for the orientation angle $\Phi\sub{K}$ but the effect may be seen in posteriors for the other parameters.\label{fig:bias}}
  \end{center}
\end{figure}

The widths calculated from multiple burst posteriors may be biased to too large values. This can happen when combining distributions of significantly different widths. When using the $k$-d tree of a distribution that is much broader than the others, a small number of leaves can contain the majority of the final posterior probability and we cannot resolve the final width. When using the $k$-d tree from a narrow distribution, there can be large leaves at the edges of the parameter ranges; because the resampled points from these leaves are uniformly distributed, they can skew the overall distribution. Since the bias increases the width, we use the smallest of the calculated values.\footnote{The distributions were checked to ensure that they did not have anomalously small widths due to a computational error.} \Figref{bias} shows an example where there is clear biasing, it shows multiple calculations of the widths alternating between using the $k$-d from the first burst posterior and the second.
In many cases the variation is comparable to the intrinsic scatter from random sampling. The $68$-percentile half-width appears more robust against biasing.

\begin{figure*}
\begin{center}
   \subfigure[{Logarithm of mass}]{\includegraphics[width=0.4\textwidth]{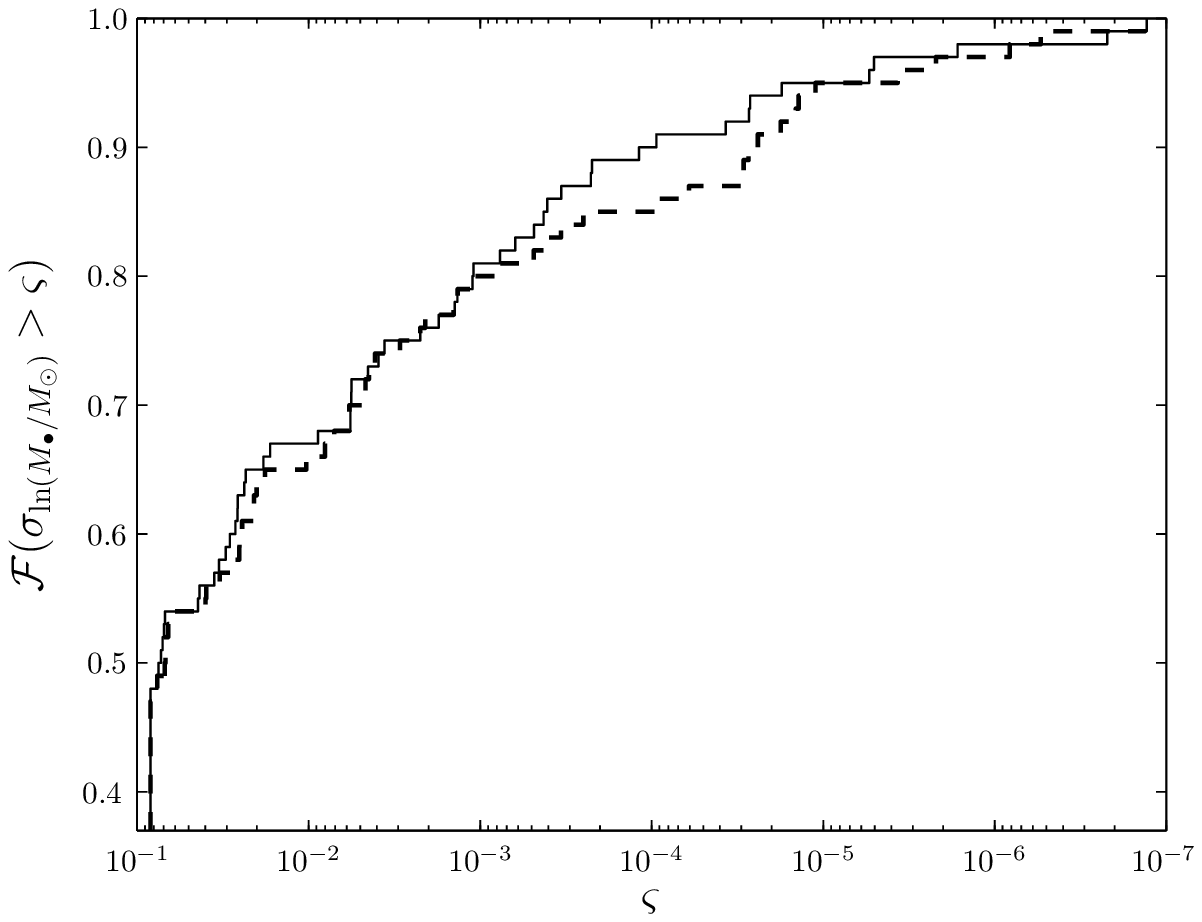}} \quad 
   \subfigure[{Spin magnitude}]{\includegraphics[width=0.4\textwidth]{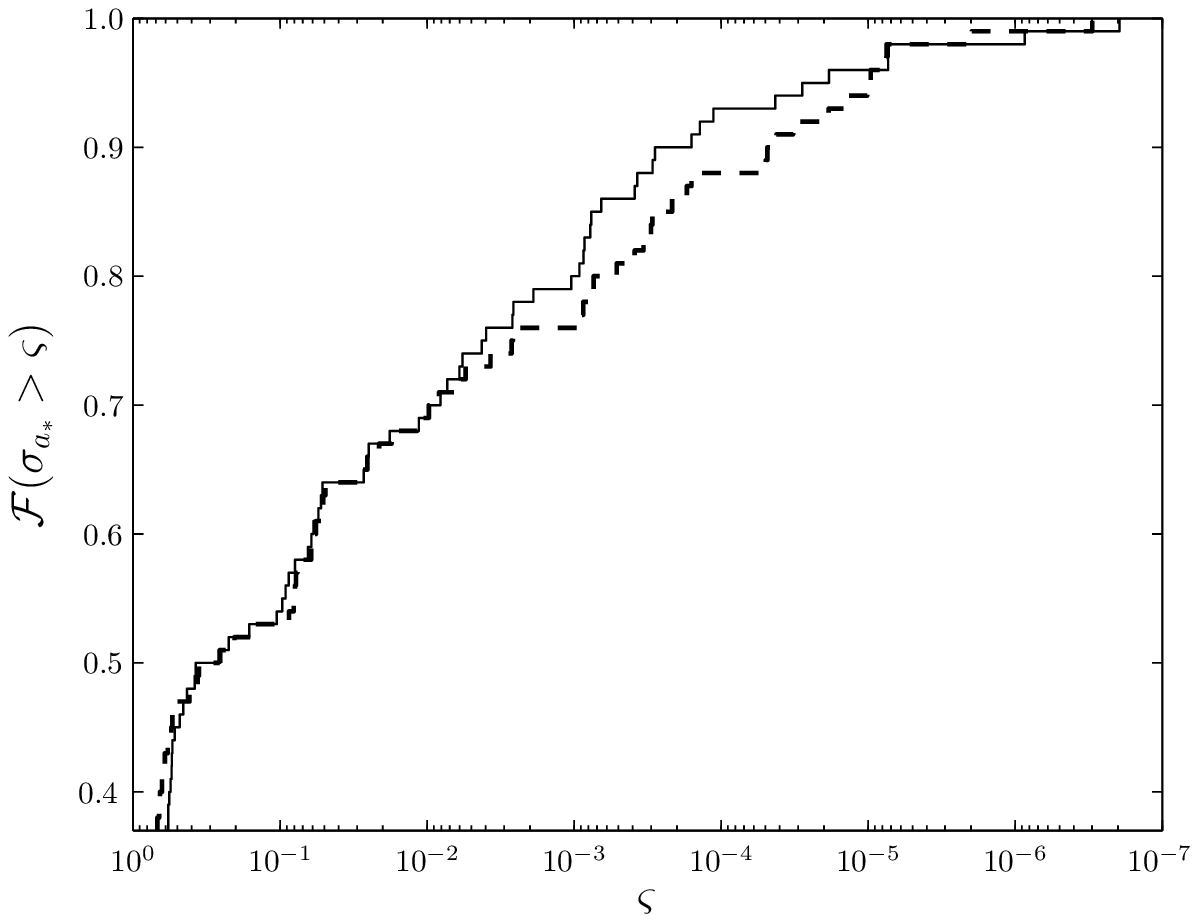}} \\
   \subfigure[{Cosine of polar angle}]{\includegraphics[width=0.4\textwidth]{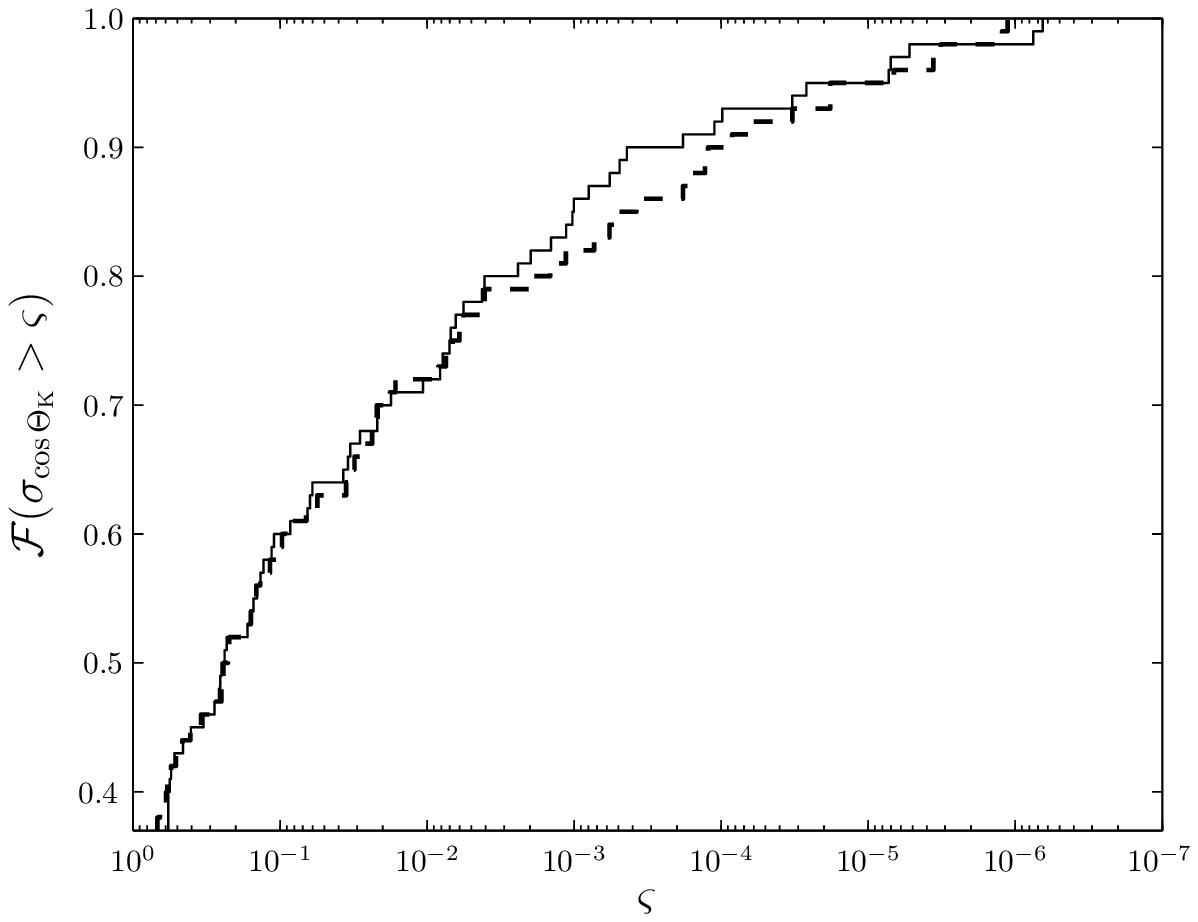}} \quad
   \subfigure[{Azimuthal angle}]{\includegraphics[width=0.4\textwidth]{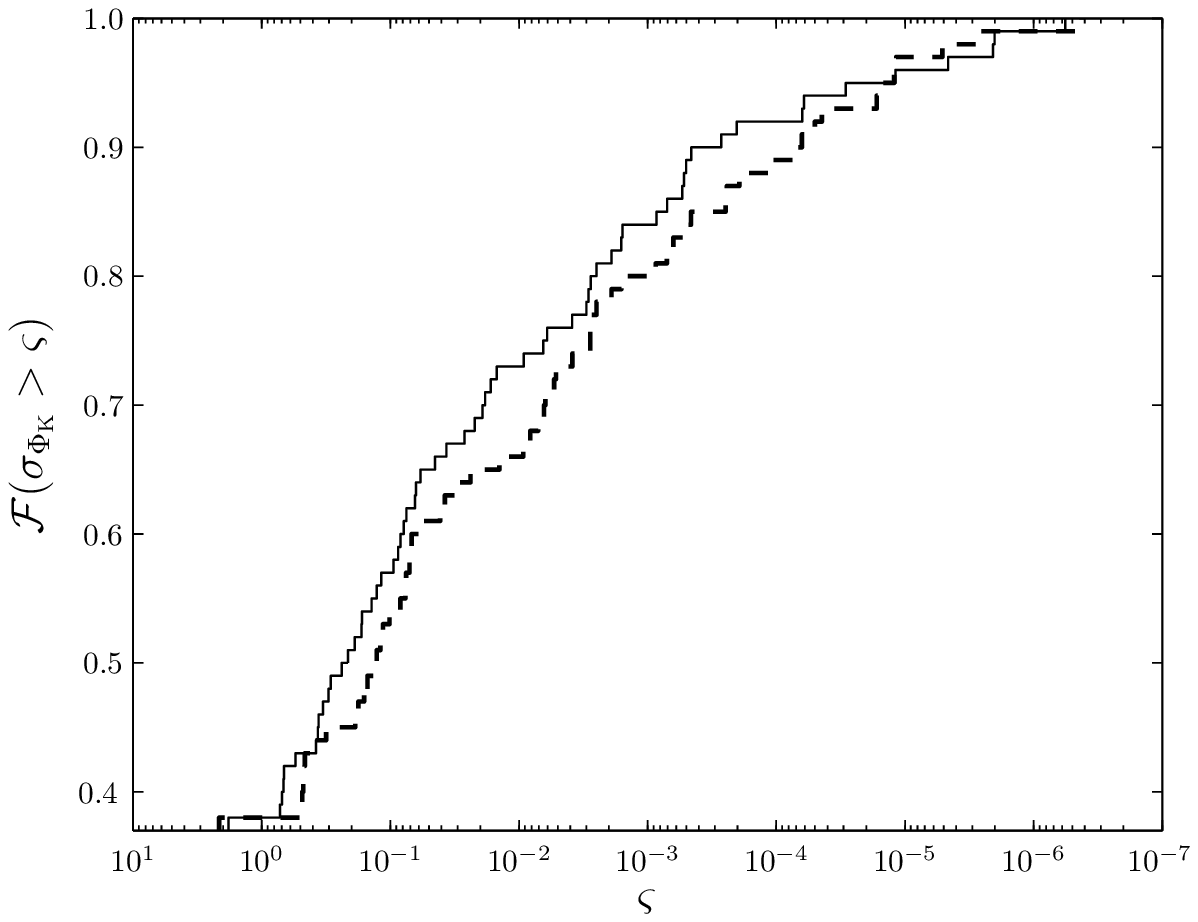}}
\caption{Cumulative proportion of mission realisations that produced posteriors with standard deviation (solid line) or $68$-percentile half-width (dashed line) larger than the abscissa value.\label{fig:Widths}}
  \end{center}
\end{figure*}

Combining the results from the set of realisations, \figref{Widths} shows the fraction of missions $\mathcal{F}(\sigma > \varsigma)$ that have posterior widths larger than $\varsigma$.
It appears that there is a $33\%$ chance of determining $\ln (M_\bullet/M_\odot)$ to a precision of $10^{-2}$ or a $10\%$ chance of determining it to $10^{-4}$. The current uncertainty of $\sim 0.08$ is bettered in over half ($52\%$) of the missions. The spin $a_\ast$ could be determined to a precision of $10^{-2}$ in $30\%$ of missions and there is a $10\%$ chance of determining it to better than $3\times 10^{-4}$.

The distribution widths work for describing parameter estimation accuracies of an individual mission; however, they are less useful for calculating an average since they are undefined when no bursts are detected. There is an alternative means of characterising what we could learn: the information entropy of the posterior distribution.

\citet{Shannon1948,Shannon1948a} introduced the idea of information entropy, which quantifies the expectation for information gained from an outcome or, equivalently, the amount of uncertainty regarding a system \citep[chapters 2 and 4]{MacKay2003}. For a discrete ensemble of probabilities $\{p_i\}$,
\begin{equation}
H(\{p_i\}) = -\sum_i p_i \ln p_i
\end{equation}
is the entropy measured in nats.\footnote{The unit is set by the base of the logarithm; the more familiar bit is calculated using base $2$, $1\units{bit} \equiv \ln(2)\units{nats}$.} This is identical to its counterpart in statistical physics up to a factor of the Boltzmann constant. Generalising from discrete to continuous probabilities is not quite as simple as exchanging the sum for an integral; it is also necessary to introduce a measure function in the logarithm, otherwise the entropy would not be invariant under a simple parameter rescaling. For a continuous probability distribution $p(\lambda)$, we work in terms of the relative entropy \citep[section 1.4]{Ihara1993}
\begin{equation}
H(p|q) = \intd{}{}{p(\lambda)\ln\left(\dfrac{p(\lambda)}{q(\lambda)}\right)}{\lambda},
\end{equation}
where $q(\lambda)$ is another probability distribution, and we have changed the sign compared to the discrete case so that the entropy is non-negative. The relative entropy, or Kullback--Leibler divergence, measures the difference between distributions and is zero only if $p(\lambda) = q(\lambda)$ everywhere; with $p(\lambda)$ as the posterior and $q(\lambda)$ as the prior, it quantifies the information gained \citep{Kullback1951}.

The relative entropy is perfect for our purpose. It is zero when we do not observe a burst or the burst is uninformative such that we do not learn anything. Otherwise it scales approximately with the (logarithm of the) posterior width, giving an indication of how much could be learnt. For example, if $p(\lambda)$ and $q(\lambda)$ were both uniform distributions, with $p(\lambda)$ having $1/z$ the width of $q(\lambda)$, $H(p|q) = \ln z$; if they were both Gaussian with equal means, and $p(\lambda)$ had $1/z$ the width of $q(\lambda)$, $H(p|q) = \ln z - (1/2)(1 - z^{-2})$.

There is one complication in using the relative entropy. We have used an improper prior for $\ln (M_\bullet/M_\odot)$; it is uniform over the entire real line and so cannot be normalised. As an alternative, we can use a Gaussian with parameters set by the current observations \citep{Gillessen2009}. The relative entropy then compares constraints from bursts with those from observing stellar motions.\footnote{Whilst this is a useful comparison, this does mean that the results are specific to the current state of knowledge and cannot be simply translated should we obtain updated measurements.}

In practice, if we were trying to infer the mass of the MBH, we would combine all our data together to form a best estimate. Then a positive entropy would indicate that the final posterior is narrower than the current observational distribution. We do not incorporate our current knowledge of the MBH mass into our prior here because we are interested in what information is contained in EMRBs alone. Therefore, our posterior from EMRBs can be broader than this observational prior. In this case the relative entropy can still be positive (since the distributions are different) even though we are not gaining information. In these cases we set the entropy to zero by hand as we have not improved our relative state of knowledge.

The entropies calculated from multiple burst posteriors may show a similar bias to the distribution widths. This would reduce the value; hence, we use the largest calculated entropy. However, the entropy appears much less sensitive to the choice of the $k$-d tree used for the multiplication than the distribution widths. The entropies corresponding to the distributions in \figref{bias} differ only by $0.04\units{nats}$, less than $1\%$ of the value.

\begin{figure*}
\begin{center}
   \subfigure[{Logarithm of mass}]{\includegraphics[width=0.4\textwidth]{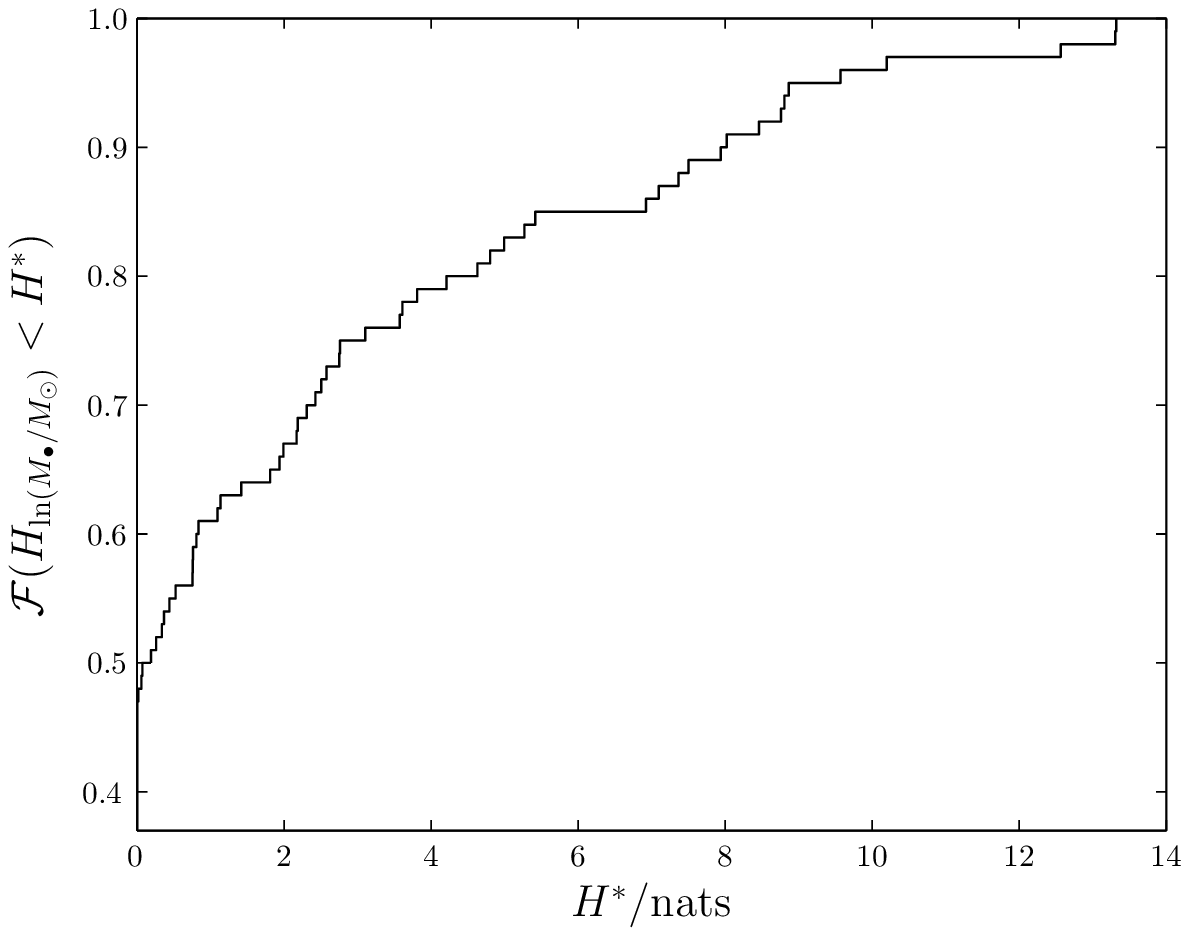}} \quad 
   \subfigure[{Spin magnitude}]{\includegraphics[width=0.4\textwidth]{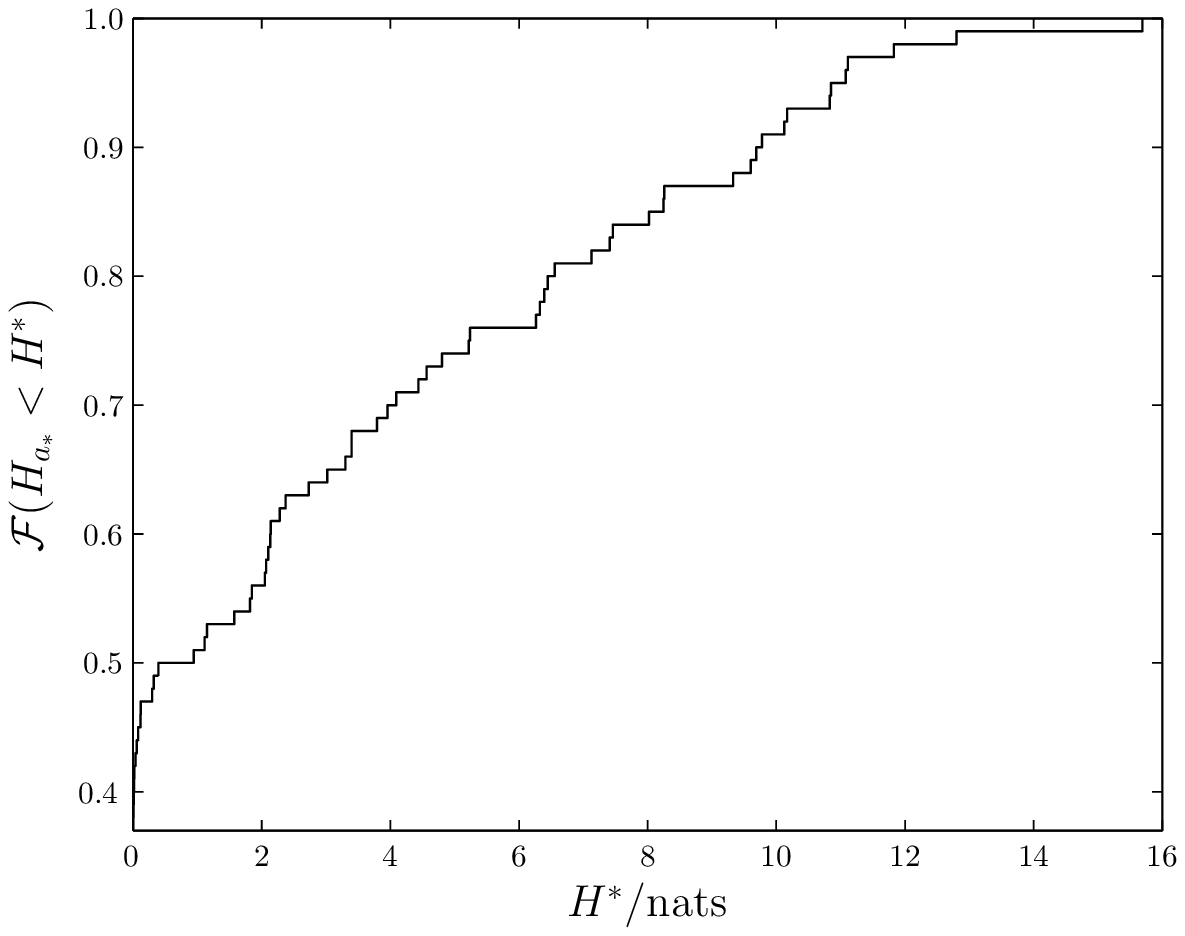}} \\
   \subfigure[{Cosine of polar angle}]{\includegraphics[width=0.4\textwidth]{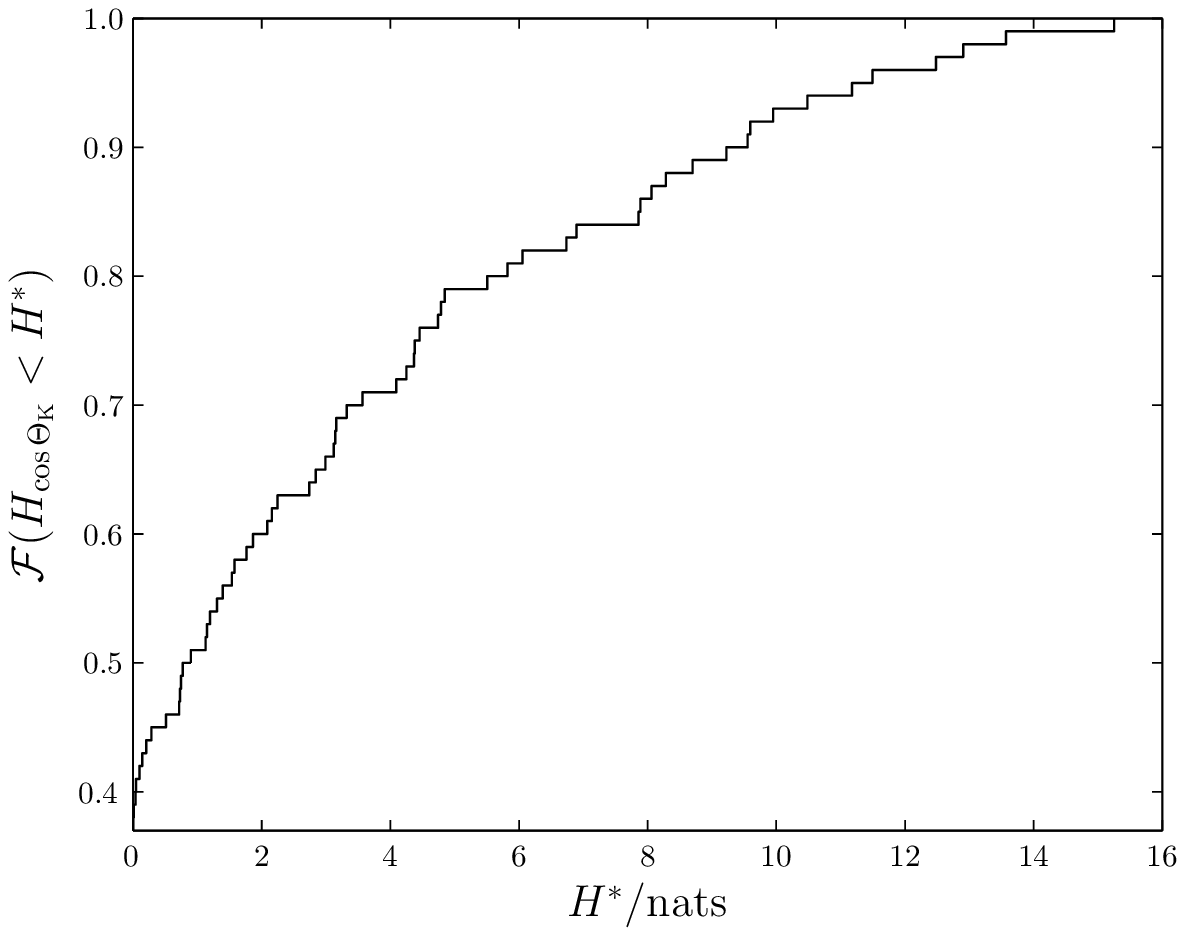}} \quad
   \subfigure[{Azimuthal angle}]{\includegraphics[width=0.4\textwidth]{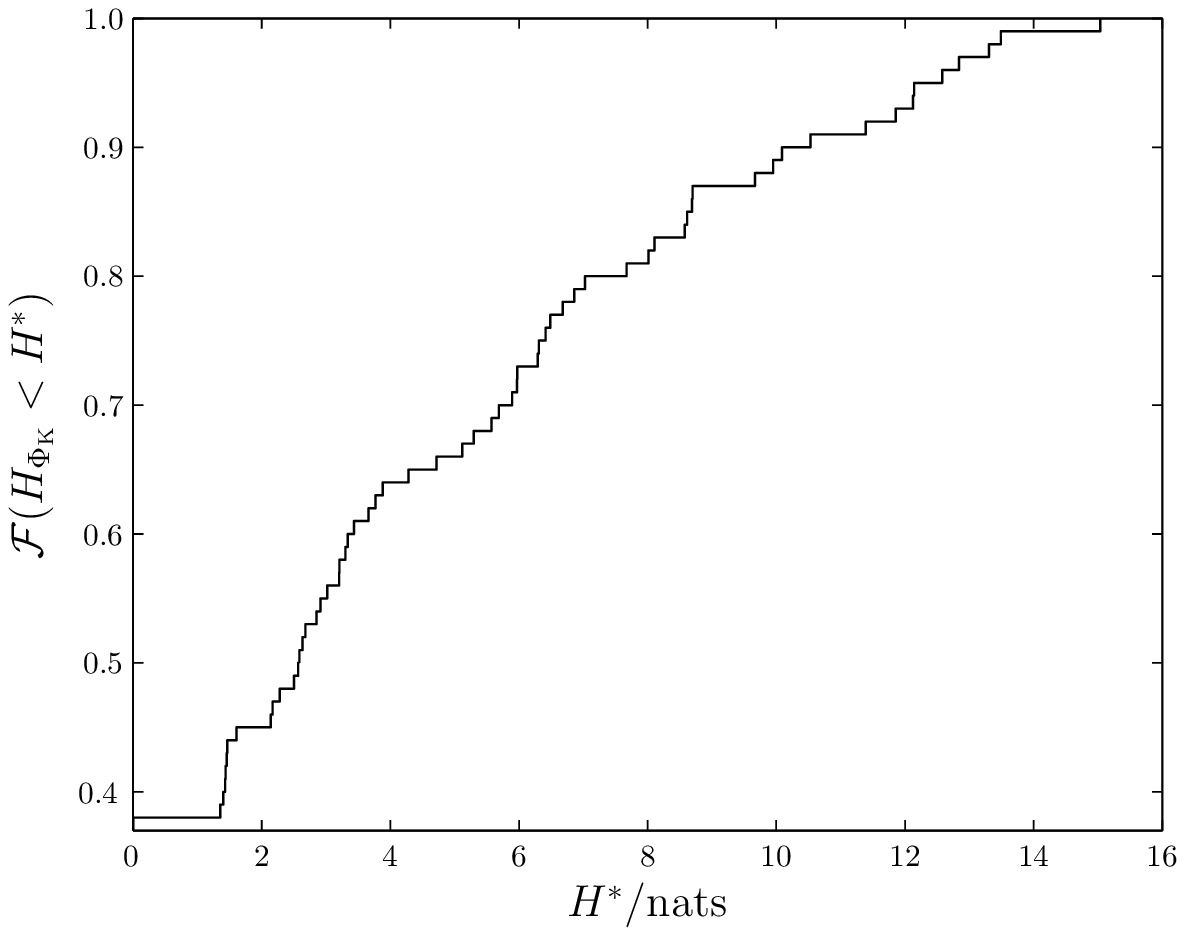}}
\caption{Cumulative proportion of mission realisations that produced posteriors with relative entropies larger than the abscissa value.\label{fig:H-ent}}
  \end{center}
\end{figure*}

The entropies are well correlated with the logarithm of the distribution widths as expected. The distributions of the fraction of missions with entropies smaller than a given value are shown in \figref{H-ent}.
They closely mirror those in \figref{Widths} (but the scale on the abscissa axis is now linear). There is a clustering at small entropies; the largest entropies are $\sim 13\units{nats}$ for $\ln(M_\bullet/M_\odot)$ and $\sim 15\units{nats}$ for the other parameters.

Taking the average across all $100$ mission realisations, we can calculate the expected information gain for each parameter:
\begin{equation}
\begin{split}
\left\langle H_{\ln(M_\bullet/M_\odot)}\right\rangle\sub{mission} & = 2.2 \pm 0.3 \units{nats}; \\
\left\langle H_{a_\ast}\right\rangle\sub{mission} & = 3.0 \pm 0.4 \units{nats}; \\
\left\langle H_{\cos\Theta\sub{K}}\right\rangle\sub{mission} & = 2.8 \pm 0.4 \units{nats}; \\
\left\langle H_{\Phi\sub{K}}\right\rangle\sub{mission} & = 3.7 \pm 0.4 \units{nats},
\end{split}
\end{equation}
using $H_\lambda \equiv H(p(\lambda)|q(\lambda))$ for brevity. The quoted uncertainties are just the standard errors calculated from the scatter of entropies and do not include any of the other uncertainties. The typical entropy is about $3 \units{nats}$; this corresponds to an improvement in the precision to which we know parameters by a factor of approximately $20$.

\section{Conclusion}\label{sec:Gal-End}

EMRBs are a potentially interesting signal for a future space-based GW detector. The most promising source for EMRBs is the GC. We built a simple theoretical model to predict the Galactic EMRB event rate. The event rate is dominated by stellar mass BHs which form a cusp about the central MBH as a consequence of mass segregation. We calculate that there could be on average $\sim1.7$ detectable bursts over a two-year mission lifetime assuming a \textit{LISA}-like interferometer, of which $\sim1.2$ are from BHs. The number of events scales linearly with the mission lifetime.

The event rate is not high: EMRBs shall not be a prolific GW source; however, the rate is not negligible. We are not guaranteed to have a burst in a mission lifetime, but it seems more likely than not that we shall have at least one.

The detectability of EMRBs is of little interest astrophysically unless we can extract information about their source systems. We investigated what could be learnt about the Galaxy's MBH. We created bursts for $100$ mission realisations and characterised the posterior probability distributions for the MBH's parameters using MCMC sampling. In a large minority ($\sim40\%$) of realisations, we cannot improve upon our existing knowledge. However, in most cases we can, and it may be possible to gain a highly precise measurement of the MBH's mass and spin.

To quantify the information gained during a mission, we used the relative entropy with respect to our current knowledge. Averaging across all the missions, we found that we can expect to gain $2.2\units{nats}$ of information about the logarithm of the mass, $3.0\units{nats}$ about the spin magnitude, $2.8\units{nats}$ about the cosine of the polar angle for the spin axis and $4.2\units{nats}$ for the azimuthal angle. The entropy scales with the logarithm of the width of the posterior distribution; hence, these entropies represent improvements in the precision of our knowledge of the parameters by factors between $\sim9$ and $\sim40$. For the mass, this would mean that the uncertainty would become $\sim1\%$; we could expect to know the spin to a precision of the order of $\sim0.1$.

It must be stressed that, whilst these results are computed accurately based on the assumptions of the model, they are only to be trusted to an order of magnitude because of the significant uncertainties in the underlying assumptions. There are a number of sources of uncertainty found throughout our analysis. First, to calculate the EMRB waveforms, we employed the NK approximation assuming parabolic trajectories. The waveforms are easy to compute, but may have inaccuracies in their amplitude profiles of a few percent at small periapses \citep{Berry2013}. This should not significantly influence detectability but may lead to differences in the shape of the posterior distributions. Since the errors in the waveforms are small, this should not qualitatively affect our results. Second, in calculating the event rate we made both mathematical and physical approximations. The former are correct to a few percent and so are negligible compared to the latter. Our model, however, does include all the relevant physical processes, and further advances the previous work of \citet{Rubbo2006} and \citet{Hopman2007}. Third, the astrophysical parameters used as inputs for our event rate calculation are themselves uncertain. Fourth, when calculating the constraints for our mission realisations, we only considered EMRBs from orbits with periapses smaller than $16 r\sub{g}$, yet whilst little information is expected from other bursts, the amount is not zero. This may lead us to slightly underestimate the total information that could be extracted from EMRBs. Finally, when combining posteriors from multiple EMRBs from the same mission, we binned our posterior distributions. This could lead to a small bias, making us underestimate the usefulness of bursts. Overall, the uncertainty in astrophysical parameters is likely to be the greatest source of error. As there are many unknowns regarding the physical assumptions it is difficult to quantify the uncertainty in our results. As we learn more about the GC, we can be more confident in our prediction.

The centre of the Galaxy is a wonderful laboratory for testing our understanding of astrophysics, in particular for learning about MBHs and their influence on their environments. EMRBs could be a new means of probing this system.

\section*{Acknowledgements}
The authors are grateful for insightful discussions with Sverre Aarseth. They authors thank Tal Alexander and Clovis Hopman for useful correspondence. They also thank Rob Cole for productive suggestions and Fabio Antonini for useful comments. CPLB is supported by STFC. JRG is supported by the Royal Society. The MCMC simulations were performed using the Darwin Supercomputer of the University of Cambridge High Performance Computing Service (\url{http://www.hpc.cam.ac.uk/}), provided by Dell Inc.\ using Strategic Research Infrastructure Funding from the Higher Education Funding Council for England.

\bibliographystyle{mn3e}
\bibliography{Galactic}

\begin{thebibliography}{127}
\expandafter\ifx\csname natexlab\endcsname\relax\def\natexlab#1{#1}\fi

\bibitem[{Alexander(2005)}]{Alexander2005}
Alexander T., 2005, Phys.\ Rep., 419, 65

\bibitem[{Alexander \& Hopman(2003)}]{Alexander2003}
Alexander T., Hopman C., 2003, ApJ, 590, L29

\bibitem[{Alexander \& Hopman(2009)}]{Alexander2009}
Alexander T., Hopman C., 2009, ApJ, 697, 1861

\bibitem[{Amaro-Seoane {et~al}\mbox{.}(2012)Amaro-Seoane, Aoudia, Babak,
  Bin\'{e}truy, Berti, Boh\'{e}, Caprini, Colpi, Cornish, Danzmann, Dufaux,
  Gair, Jennrich, Jetzer, Klein, Lang, Lobo, Littenberg, McWilliams, Nelemans,
  Petiteau, Porter, Schutz, Sesana, Stebbins, Sumner, Vallisneri, Vitale,
  Volonteri, \& Ward}]{Amaro-Seoane2012a}
Amaro-Seoane P. {et~al.}, 2012, Class.\ Quantum Grav., 29, 124016

\bibitem[{Amaro-Seoane {et~al}\mbox{.}(2007)Amaro-Seoane, Gair, Freitag,
  Miller, Mandel, Cutler, \& Babak}]{Amaro-Seoane2007}
Amaro-Seoane P., Gair J.~R., Freitag M., Miller M.~C., Mandel I., Cutler C.~J.,
  Babak S., 2007, Class.\ Quantum Grav., 24, R113

\bibitem[{Amaro-Seoane \& Preto(2011)}]{Amaro-Seoane2011}
Amaro-Seoane P., Preto M., 2011, Class.\ Quantum Grav., 28, 094017

\bibitem[{Andrieu \& Thoms(2008)}]{Andrieu2008}
Andrieu C., Thoms J., 2008, Stat.\ Comp., 18, 343

\bibitem[{Antonini(2013)}]{Antonini2013}
Antonini F., 2013, ApJ, 763, 62

\bibitem[{Antonini {et~al}\mbox{.}(2012)Antonini, Capuzzo-Dolcetta,
  Mastrobuono-Battisti, \& Merritt}]{Antonini2011a}
Antonini F., Capuzzo-Dolcetta R., Mastrobuono-Battisti A., Merritt D., 2012,
  ApJ, 750, 111

\bibitem[{Antonini \& Merritt(2012)}]{Antonini2011}
Antonini F., Merritt D., 2012, ApJ, 745, 83

\bibitem[{Antonucci {et~al}\mbox{.}(2012)Antonucci, Armano, Audley, Auger,
  Benedetti, Binetruy, Bogenstahl, Bortoluzzi, Bosetti, Brandt, Caleno,
  Ca\~{n}izares, Cavalleri, Cesa, Chmeissani, Conchillo, Congedo, Cristofolini,
  Cruise, Danzmann, {De Marchi}, Diaz-Aguilo, Diepholz, Dixon, Dolesi, Dunbar,
  Fauste, Ferraioli, Ferrone, Fichter, Fitzsimons, Freschi, Marin,
  Marirrodriga, Gerndt, Gesa, Gilbert, Giardini, Grimani, Grynagier, Guillaume,
  Guzm\'{a}n, Harrison, Heinzel, Hern\'{a}ndez, Hewitson, Hollington, Hough,
  Hoyland, Hueller, Huesler, Jennrich, Jetzer, Johlander, Karnesis, Killow,
  Llamas, Lloro, Lobo, Maarschalkerweerd, Madden, Mance, Mateos, McNamara,
  Mendes, Mitchell, Monsky, Nicolini, Nicolodi, Nofrarias, Pedersen,
  Perreur-Lloyd, Plagnol, Prat, Racca, Ramos-Castro, Reiche, Perez, Robertson,
  Rozemeijer, Sanjuan, Schleicher, Schulte, Shaul, Stagnaro, Strandmoe, Steier,
  Sumner, Taylor, Texier, Trenkel, Tu, Vitale, Wanner, Ward, Waschke, Wass,
  Weber, Ziegler, \& Zweifel}]{Antonucci2012}
Antonucci F. {et~al.}, 2012, Class.\ Quantum Grav., 29, 124014

\bibitem[{Anza {et~al}\mbox{.}(2005)Anza, Armano, Balaguer, Benedetti,
  Boatella, Bosetti, Bortoluzzi, Brandt, Braxmaier, Caldwell, Carbone,
  Cavalleri, Ciccolella, Cristofolini, Cruise, Lio, Danzmann, Desiderio,
  Dolesi, Dunbar, Fichter, Garcia, Garcia-Berro, Marin, Gerndt, Gianolio,
  Giardini, Gruenagel, Hammesfahr, Heinzel, Hough, Hoyland, Hueller, Jennrich,
  Johann, Kemble, Killow, Kolbe, Landgraf, Lobo, Lorizzo, Mance, Middleton,
  Nappo, Nofrarias, Racca, Ramos, Robertson, Sallusti, Sandford, Sanjuan,
  Sarra, Selig, Shaul, Smart, Smit, Stagnaro, Sumner, Tirabassi, Tobin, Vitale,
  Wand, Ward, Weber, \& Zweifel}]{Anza2005}
Anza S. {et~al.}, 2005, Class.\ Quantum Grav., 22, S125

\bibitem[{Babak {et~al}\mbox{.}(2007)Babak, Fang, Gair, Glampedakis, \&
  Hughes}]{Babak2007}
Babak S., Fang H., Gair J., Glampedakis K., Hughes S., 2007, Phys.\ Rev.\ D,
  75, 024005

\bibitem[{Bahcall \& Wolf(1976)}]{Bahcall1976}
Bahcall J.~N., Wolf R.~A., 1976, ApJ, 209, 214

\bibitem[{Bahcall \& Wolf(1977)}]{Bahcall1977}
Bahcall J.~N., Wolf R.~A., 1977, ApJ, 216, 883

\bibitem[{Bar-Or {et~al}\mbox{.}(2013)Bar-Or, Kupi, \& Alexander}]{Bar-Or2012}
Bar-Or B., Kupi G., Alexander T., 2013, ApJ, 764, 52

\bibitem[{Barack(2009)}]{Barack2009}
Barack L., 2009, Class.\ Quantum Grav., 26, 213001(56)

\bibitem[{Barack \& Cutler(2004)}]{Barack2004}
Barack L., Cutler C., 2004, Phys.\ Rev.\ D, 69, 082005

\bibitem[{Bartos {et~al}\mbox{.}(2013)Bartos, Haiman, Kocsis, \&
  M{\'a}rka}]{Bartos2013}
Bartos I., Haiman Z., Kocsis B., M{\'a}rka S., 2013, Phys.\ Rev.\ Lett., 110,
  221102

\bibitem[{Baumgardt {et~al}\mbox{.}(2004)Baumgardt, Makino, \&
  Ebisuzaki}]{Baumgardt2004}
Baumgardt H., Makino J., Ebisuzaki T., 2004, ApJ, 613, 1133

\bibitem[{Bekenstein(1973)}]{Bekenstein1973}
Bekenstein J.~D., 1973, ApJ, 183, 657

\bibitem[{Bekenstein \& Maoz(1992)}]{Bekenstein1992}
Bekenstein J.~D., Maoz E., 1992, ApJ, 390, 79

\bibitem[{Bender {et~al}\mbox{.}(1998)Bender, Brillet, Ciufolini, Cruise,
  Cutler, Danzmann, Fidecaro, Folkner, Hough, McNamara, Peterseim, Robertson,
  Rodrigues, R\"{u}diger, Sandford, Sch\"{a}fer, Schilling, Schutz, Speake,
  Stebbins, Sumner, Touboul, Vinet, Vitale, Ward, \& Winkler}]{Bender1998}
Bender P. {et~al.}, 1998, {LISA Pre-Phase A Report}. Tech. rep.,
  Max-Planck-Institut f\"{u}r Quantenoptik, Garching

\bibitem[{Berry \& Gair(2013{\natexlab{a}})}]{Berry2013a}
Berry C. P.~L., Gair J.~R., 2013{\natexlab{a}}, MNRAS, 433, 3572

\bibitem[{Berry \& Gair(2013{\natexlab{b}})}]{Berry2013}
Berry C. P.~L., Gair J.~R., 2013{\natexlab{b}}, MNRAS, 429, 589

\bibitem[{Binney \& Tremaine(2008)}]{Binney2008}
Binney J., Tremaine S., 2008, {Galactic Dynamics}, 2nd edn., Princeton Series
  in Astrophysics. Princeton Univ.\ Press, Princeton, NJ

\bibitem[{Chandrasekhar(1941{\natexlab{a}})}]{Chandrasekhar1941a}
Chandrasekhar S., 1941{\natexlab{a}}, ApJ, 94, 511

\bibitem[{Chandrasekhar(1941{\natexlab{b}})}]{Chandrasekhar1941}
Chandrasekhar S., 1941{\natexlab{b}}, ApJ, 93, 285

\bibitem[{Chandrasekhar(1943{\natexlab{a}})}]{Chandrasekhar1943a}
Chandrasekhar S., 1943{\natexlab{a}}, ApJ, 97, 255

\bibitem[{Chandrasekhar(1943{\natexlab{b}})}]{Chandrasekhar1943}
Chandrasekhar S., 1943{\natexlab{b}}, ApJ, 98, 54

\bibitem[{Chandrasekhar(1960)}]{Chandrasekhar1960}
Chandrasekhar S., 1960, {Principles of Stellar Dynamics}, enlarged edn. Dover
  Publications, New York

\bibitem[{Chatterjee {et~al}\mbox{.}(2002)Chatterjee, Hernquist, \&
  Loeb}]{Chatterjee2002}
Chatterjee P., Hernquist L., Loeb A., 2002, ApJ, 572, 371

\bibitem[{Cohn \& Kulsrud(1978)}]{Cohn1978}
Cohn H., Kulsrud R.~M., 1978, ApJ, 226, 1087

\bibitem[{Cutler(1998)}]{Cutler1998}
Cutler C., 1998, Phys.\ Rev.\ D, 57, 7089

\bibitem[{Cutler \& Flanagan(1994)}]{Cutler1994}
Cutler C., Flanagan E.~E., 1994, Phys.\ Rev.\ D, 49, 2658

\bibitem[{Danzmann \& R\"{u}diger(2003)}]{Danzmann2003}
Danzmann K., R\"{u}diger A., 2003, Class.\ Quantum Grav., 20, S1

\bibitem[{de~Berg {et~al}\mbox{.}(2008)de~Berg, Cheong, van Kreveld, \&
  Overmars}]{Berg2008}
de~Berg M., Cheong O., van Kreveld M., Overmars M., 2008, {Computational
  Geometry: Algorithms and Applications}, 3rd edn. Springer, Berlin

\bibitem[{Doeleman {et~al}\mbox{.}(2008)Doeleman, Weintroub, Rogers, Plambeck,
  Freund, Tilanus, Friberg, Ziurys, Moran, Corey, Young, Smythe, Titus,
  Marrone, Cappallo, Bock, Bower, Chamberlin, Davis, Krichbaum, Lamb, Maness,
  Niell, Roy, Strittmatter, Werthimer, Whitney, \& Woody}]{Doeleman2008}
Doeleman S.~S. {et~al.}, 2008, Nat, 455, 78

\bibitem[{Drasco \& Hughes(2004)}]{Drasco2004}
Drasco S., Hughes S., 2004, Phys.\ Rev.\ D, 69, 044015

\bibitem[{Eilon {et~al}\mbox{.}(2009)Eilon, Kupi, \& Alexander}]{Eilon2009}
Eilon E., Kupi G., Alexander T., 2009, ApJ, 698, 641

\bibitem[{Ferrarese \& Ford(2005)}]{Ferrarese2005}
Ferrarese L., Ford H., 2005, Space Sci.\ Rev., 116, 523

\bibitem[{Frank \& Rees(1976)}]{Frank1976}
Frank J., Rees M.~J., 1976, MNRAS, 176, 633

\bibitem[{Freitag {et~al}\mbox{.}(2006)Freitag, Amaro-Seoane, \&
  Kalogera}]{Freitag2006}
Freitag M., Amaro-Seoane P., Kalogera V., 2006, ApJ, 649, 91

\bibitem[{Freitag \& Benz(2005)}]{Freitag2005}
Freitag M., Benz W., 2005, MNRAS, 358, 1133

\bibitem[{Gair {et~al}\mbox{.}(2005)Gair, Kennefick, \& Larson}]{Gair2005}
Gair J.~R., Kennefick D.~J., Larson S.~L., 2005, Phys.\ Rev.\ D, 72, 084009

\bibitem[{Genzel {et~al}\mbox{.}(2010)Genzel, Eisenhauer, \&
  Gillessen}]{Genzel2010}
Genzel R., Eisenhauer F., Gillessen S., 2010, Rev.\ Mod.\ Phys., 82, 3121

\bibitem[{Genzel {et~al}\mbox{.}(2003)Genzel, Schoedel, Ott, Eisenhauer,
  Hofmann, Lehnert, Eckart, Alexander, Sternberg, Lenzen, Clenet, Lacombe,
  Rouan, Renzini, \& Tacconi-Garman}]{Genzel2003}
Genzel R. {et~al.}, 2003, ApJ, 594, 812

\bibitem[{Ghez {et~al}\mbox{.}(2008)Ghez, Salim, Weinberg, Lu, Do, Dunn,
  Matthews, Morris, Yelda, Becklin, Kremenek, Milosavljevic, \&
  Naiman}]{Ghez2008}
Ghez A.~M. {et~al.}, 2008, ApJ, 689, 1044

\bibitem[{Gillessen {et~al}\mbox{.}(2009)Gillessen, Eisenhauer, Trippe,
  Alexander, Genzel, Martins, \& Ott}]{Gillessen2009}
Gillessen S., Eisenhauer F., Trippe S., Alexander T., Genzel R., Martins F.,
  Ott T., 2009, ApJ, 692, 1075

\bibitem[{Glampedakis(2005)}]{Glampedakis2005}
Glampedakis K., 2005, Class.\ Quantum Grav., 22, S605

\bibitem[{Gualandris \& Merritt(2012)}]{Gualandris2012}
Gualandris A., Merritt D., 2012, ApJ, 744, 74

\bibitem[{G\"{u}rkan \& Hopman(2007)}]{Gurkan2007}
G\"{u}rkan M.~A., Hopman C., 2007, MNRAS, 379, 1083

\bibitem[{Haario {et~al}\mbox{.}(1999)Haario, Saksman, \&
  Tamminen}]{Haario1999}
Haario H., Saksman E., Tamminen J., 1999, Comp.\ Stat., 14, 375

\bibitem[{Hastings(1970)}]{Hastings1970}
Hastings W.~K., 1970, Biometrika, 57, 97

\bibitem[{Hills(1975)}]{Hills1975}
Hills J.~G., 1975, Nat, 254, 295

\bibitem[{Hobson {et~al}\mbox{.}(2006)Hobson, Efstathiou, \&
  Lasenby}]{Hobson2006}
Hobson M.~P., Efstathiou G., Lasenby A., 2006, {General Relativity: An
  Introduction for Physicists}. Cambridge Univ.\ Press, Cambridge

\bibitem[{Hopman(2009)}]{Hopman2009}
Hopman C., 2009, ApJ, 700, 1933

\bibitem[{Hopman \& Alexander(2005)}]{Hopman2005}
Hopman C., Alexander T., 2005, ApJ, 629, 362

\bibitem[{Hopman \& Alexander(2006{\natexlab{a}})}]{Hopman2006}
Hopman C., Alexander T., 2006{\natexlab{a}}, ApJ, 645, 1152

\bibitem[{Hopman \& Alexander(2006{\natexlab{b}})}]{Hopman2006a}
Hopman C., Alexander T., 2006{\natexlab{b}}, ApJ, 645, L133

\bibitem[{Hopman {et~al}\mbox{.}(2007)Hopman, Freitag, \& Larson}]{Hopman2007}
Hopman C., Freitag M., Larson S.~L., 2007, MNRAS, 378, 129

\bibitem[{Ihara(1993)}]{Ihara1993}
Ihara S., 1993, {Information Theory for Continuous Systems}. World Scientific
  Publishing Company, London

\bibitem[{Jennrich {et~al}\mbox{.}(2011)Jennrich, Binetruy, Colpi, Danzmann,
  Jetzer, Lobo, Nelemans, Schutz, Stebbins, Sumner, Vitale, \&
  Ward}]{Jennrich2011}
Jennrich O. {et~al.}, 2011, {NGO Revealing a hidden Universe: opening a new
  chapter of discovery}. Tech. rep., ESA, Noordwijk

\bibitem[{Just {et~al}\mbox{.}(2011)Just, Khan, Berczik, Ernst, \&
  Spurzem}]{Just2011}
Just A., Khan F.~M., Berczik P., Ernst A., Spurzem R., 2011, MNRAS, 411, 653

\bibitem[{Kesden(2012)}]{Kesden2012}
Kesden M., 2012, Phys.\ Rev.\ D, 85, 024037

\bibitem[{Keshet {et~al}\mbox{.}(2009)Keshet, Hopman, \&
  Alexander}]{Keshet2009}
Keshet U., Hopman C., Alexander T., 2009, ApJ, 698, L64

\bibitem[{Kobayashi {et~al}\mbox{.}(2004)Kobayashi, Laguna, Phinney, \&
  Meszaros}]{Kobayashi2004}
Kobayashi S., Laguna P., Phinney E.~S., Meszaros P., 2004, ApJ, 615, 855

\bibitem[{Kullback \& Leibler(1951)}]{Kullback1951}
Kullback S., Leibler R.~A., 1951, Ann.\ Math.\ Stat., 22, 79

\bibitem[{Landau \& Lifshitz(1958)}]{Landau1958}
Landau L.~D., Lifshitz E.~M., 1958, {Statistical Physics}, Course of
  Theoretical Physics. Pergamon Press, London

\bibitem[{Li {et~al}\mbox{.}(2012)Li, Liu, Berczik, Chen, \& Spurzem}]{Li2012}
Li S., Liu F.~K., Berczik P., Chen X., Spurzem R., 2012, ApJ, 748, 65

\bibitem[{Lightman \& Shapiro(1977)}]{Lightman1977}
Lightman A.~P., Shapiro S.~L., 1977, ApJ, 211, 244

\bibitem[{Lynden-Bell(1967)}]{Lynden-Bell1967}
Lynden-Bell D., 1967, MNRAS, 136, 101

\bibitem[{Lynden-Bell(1969)}]{Lynden-Bell1969}
Lynden-Bell D., 1969, Nat, 223, 690

\bibitem[{Lynden-Bell \& Kalnajs(1972)}]{Lynden-Bell1972}
Lynden-Bell D., Kalnajs A.~J., 1972, MNRAS, 157, 1

\bibitem[{MacKay(2003)}]{MacKay2003}
MacKay D. J.~C., 2003, {Information Theory, Inference and Learning Algorithms}.
  Cambridge Univ.\ Press, Cambridge

\bibitem[{Madigan {et~al}\mbox{.}(2011)Madigan, Hopman, \& Levin}]{Madigan2011}
Madigan A.-M., Hopman C., Levin Y., 2011, ApJ, 738, 99

\bibitem[{Maoz(1993)}]{Maoz1993}
Maoz E., 1993, MNRAS, 263, 75

\bibitem[{Merritt(2010)}]{Merritt2010a}
Merritt D., 2010, ApJ, 718, 739

\bibitem[{Merritt {et~al}\mbox{.}(2011)Merritt, Alexander, Mikkola, \&
  Will}]{Merritt2011}
Merritt D., Alexander T., Mikkola S., Will C.~M., 2011, Phys.\ Rev.\ D, 84,
  044024

\bibitem[{Merritt {et~al}\mbox{.}(2007)Merritt, Berczik, \& Laun}]{Merritt2007}
Merritt D., Berczik P., Laun F., 2007, AJ, 133, 553

\bibitem[{Metropolis {et~al}\mbox{.}(1953)Metropolis, Rosenbluth, Rosenbluth,
  Teller, \& Teller}]{Metropolis1953}
Metropolis N., Rosenbluth A.~W., Rosenbluth M.~N., Teller A.~H., Teller E.,
  1953, J.\ Chem.\ Phys., 21, 1087

\bibitem[{Meyer {et~al}\mbox{.}(2012)Meyer, Ghez, Sch{\"o}del, Yelda, Boehle,
  Lu, Do, Morris, Becklin, \& Matthews}]{Meyer2012}
Meyer L. {et~al.}, 2012, Sci, 338, 84

\bibitem[{Miralda-Escude \& Gould(2000)}]{MiraldaEscude2000}
Miralda-Escude J., Gould A., 2000, ApJ, 545, 847

\bibitem[{Misner {et~al}\mbox{.}(1973)Misner, Thorne, \& Wheeler}]{Misner1973}
Misner C.~W., Thorne K.~S., Wheeler J.~A., 1973, {Gravitation}. Freeman, New
  York

\bibitem[{Mulder(1983)}]{Mulder1983}
Mulder W.~A., 1983, A\&{}A, 117, 9

\bibitem[{Murphy {et~al}\mbox{.}(1991)Murphy, Cohn, \& Durisen}]{Murphy1991}
Murphy B.~W., Cohn H.~N., Durisen R.~H., 1991, ApJ, 370, 60

\bibitem[{Nelson \& Tremaine(1999)}]{Nelson1999}
Nelson R.~W., Tremaine S., 1999, MNRAS, 306, 1

\bibitem[{O'Leary {et~al}\mbox{.}(2009)O'Leary, Kocsis, \& Loeb}]{O'Leary2009}
O'Leary R.~M., Kocsis B., Loeb A., 2009, MNRAS, 395, 2127

\bibitem[{Olver {et~al}\mbox{.}(2010)Olver, Lozier, Boisvert, \&
  Clark}]{Olver2010}
Olver F. W.~J., Lozier Daniel W., Boisvert R.~F., Clark C.~W., eds., 2010,
  {NIST Handbook of Mathematical Functions}. Cambridge Univ.\ Press, Cambridge

\bibitem[{Peebles(1972)}]{Peebles1972}
Peebles P. J.~E., 1972, ApJ, 178, 371

\bibitem[{Peters(1964)}]{Peters1964}
Peters P.~C., 1964, Physical Review, 136, B1224

\bibitem[{Peters \& Mathews(1963)}]{Peters1963}
Peters P.~C., Mathews J., 1963, Physical Review, 131, 435

\bibitem[{Press(1977)}]{Press1977}
Press W., 1977, Phys.\ Rev.\ D, 15, 965

\bibitem[{Preto \& Amaro-Seoane(2010)}]{Preto2010}
Preto M., Amaro-Seoane P., 2010, ApJ, 708, L42

\bibitem[{Rauch \& Ingalls(1998)}]{Rauch1998}
Rauch K.~P., Ingalls B., 1998, MNRAS, 299, 1231

\bibitem[{Rauch \& Tremaine(1996)}]{Rauch1996}
Rauch K.~P., Tremaine S., 1996, New Astron., 1, 149

\bibitem[{Rees(1984)}]{Rees1984}
Rees M.~J., 1984, ARA\&{}A, 22, 471

\bibitem[{Rees(1988)}]{Rees1988}
Rees M.~J., 1988, Nat, 333, 523

\bibitem[{Reid \& Brunthaler(2004)}]{Reid2004}
Reid M.~J., Brunthaler A., 2004, ApJ, 616, 872

\bibitem[{Reid {et~al}\mbox{.}(2003)Reid, Menten, Genzel, Ott, Sch\"{o}del, \&
  Brunthaler}]{Reid2003}
Reid M.~J., Menten K.~M., Genzel R., Ott T., Sch\"{o}del R., Brunthaler A.,
  2003, Astron.\ Nach., 324, 505

\bibitem[{Reid {et~al}\mbox{.}(1999)Reid, Readhead, Vermeulen, \&
  Treuhaft}]{Reid1999}
Reid M.~J., Readhead A. C.~S., Vermeulen R.~C., Treuhaft R.~N., 1999, ApJ, 524,
  816

\bibitem[{Roberts \& Rosenthal(2007)}]{Roberts2007}
Roberts G.~O., Rosenthal J.~S., 2007, J.\ Appl.\ Prob., 44, 458

\bibitem[{Rubbo {et~al}\mbox{.}(2006)Rubbo, Holley-Bockelmann, \&
  Finn}]{Rubbo2006}
Rubbo L.~J., Holley-Bockelmann K., Finn L.~S., 2006, ApJ, 649, L25

\bibitem[{Ruffini \& Sasaki(1981)}]{Ruffini1981}
Ruffini R., Sasaki M., 1981, Prog.\ Theor.\ Phys., 66, 1627

\bibitem[{Sch\"{o}del {et~al}\mbox{.}(2007)Sch\"{o}del, Eckart, Alexander,
  Merritt, Genzel, Sternberg, Meyer, Kul, Moultaka, Ott, \&
  Straubmeier}]{Schodel2007}
Sch\"{o}del R. {et~al.}, 2007, A\&{}A, 469, 125

\bibitem[{Shannon(1948{\natexlab{a}})}]{Shannon1948}
Shannon C.~E., 1948{\natexlab{a}}, Bell Syst.\ Tech.\ J., 27, 379

\bibitem[{Shannon(1948{\natexlab{b}})}]{Shannon1948a}
Shannon C.~E., 1948{\natexlab{b}}, Bell Syst.\ Tech.\ J., 27, 623

\bibitem[{Shapiro \& Marchant(1978)}]{Shapiro1978}
Shapiro S.~L., Marchant A.~B., 1978, ApJ, 225, 603

\bibitem[{Sidery {et~al}\mbox{.}(2013)Sidery, Gair, \& Mandel}]{Sidery2013}
Sidery T., Gair J.~R., Mandel I., 2013, in preparation

\bibitem[{Sigurdsson \& Rees(1997)}]{Sigurdsson1997}
Sigurdsson S., Rees M.~J., 1997, MNRAS, 284, 318

\bibitem[{Spitzer \& Hart(1971)}]{Spitzer1971}
Spitzer L., Hart M.~H., 1971, ApJ, 164, 399

\bibitem[{Spitzer \& Shapiro(1972)}]{Spitzer1972}
Spitzer L., Shapiro S.~L., 1972, ApJ, 173, 529

\bibitem[{Spitzer(1987)}]{Spitzer1987}
Spitzer, Jr. L., 1987, {Dynamical Evolution of Globular Clusters}, Princeton
  Series in Astrophysics. Princeton Univ.\ Press, Princeton, NJ

\bibitem[{Spitzer \& Harm(1958)}]{Spitzer1958}
Spitzer, Jr. L., Harm R., 1958, ApJ, 127, 544

\bibitem[{Spitzer \& Schwarzschild(1951)}]{Spitzer1951}
Spitzer, Jr. L., Schwarzschild M., 1951, ApJ, 114, 385

\bibitem[{Tanaka {et~al}\mbox{.}(1993)Tanaka, Shibata, Sasaki, Tagoshi, \&
  Nakamura}]{Tanaka1993}
Tanaka T., Shibata M., Sasaki M., Tagoshi H., Nakamura T., 1993, Prog.\ Theor.\
  Phys., 90, 65

\bibitem[{Toonen {et~al}\mbox{.}(2009)Toonen, Hopman, \& Freitag}]{Toonen2009}
Toonen S., Hopman C., Freitag M., 2009, MNRAS, 398, 1228

\bibitem[{Tremaine {et~al}\mbox{.}(2002)Tremaine, Gebhardt, Bender, Bower,
  Dressler, Faber, Filippenko, Green, Grillmair, Ho, Kormendy, Lauer,
  Magorrian, Pinkney, \& Richstone}]{Tremaine2002}
Tremaine S. {et~al.}, 2002, ApJ, 574, 740

\bibitem[{Tremaine \& Weinberg(1984)}]{Tremaine1984}
Tremaine S., Weinberg M.~D., 1984, MNRAS, 209, 729

\bibitem[{Turner(1977)}]{Turner1977}
Turner M., 1977, ApJ, 216, 610

\bibitem[{Volonteri(2010)}]{Volonteri2010}
Volonteri M., 2010, A\&{}AR, 18, 279

\bibitem[{Weinberg(1986)}]{Weinberg1986}
Weinberg M.~D., 1986, ApJ, 300, 93

\bibitem[{Weinberg(2012)}]{Weinberg2012}
Weinberg M.~D., 2012, Bayesian Anal., 7, 737

\bibitem[{Wyse(2008)}]{Wyse2008}
Wyse R. F.~G., 2008, in ASP Conference Series, Vol. 399, Panoramic Views of
  Galaxy Formation and Evolution, Kodama T., Yamada T., Aoki K., eds., Astron.\
  Soc.\ Pac., San Francisco, pp. 445--448

\bibitem[{Young(1977)}]{Young1977}
Young P.~J., 1977, ApJ, 217, 287

\bibitem[{Yunes {et~al}\mbox{.}(2008)Yunes, Sopuerta, Rubbo, \&
  Holley-Bockelmann}]{Yunes2008}
Yunes N.~N., Sopuerta C.~F., Rubbo L.~J., Holley-Bockelmann K., 2008, ApJ, 675,
  604

\bibitem[{Yusef-Zadeh {et~al}\mbox{.}(1999)Yusef-Zadeh, Choate, \&
  Cotton}]{Yusef-Zadeh1999}
Yusef-Zadeh F., Choate D., Cotton W., 1999, ApJ, 518, L33

\end{thebibliography}

\appendix

\begin{onecolumn}

\section{Chandrasekhar's relaxation time-scale}\label{sec:time-scale}

\citet[chapter 2]{Chandrasekhar1960} defined a relaxation time-scale for a stellar system by approximating the fluctuations in the stellar gravitational potential as a series of two-body encounters. The time over which the squared change in energy is equal to the squared (initial) kinetic energy of the star is the time taken for relaxation. Relaxation is mediated by dynamical friction (\citealt{Chandrasekhar1943a}; \citealt[section 1.2]{Binney2008}). This can be understood as the drag induced on a star by the overdensity of field stars deflected by its passage \citep{Mulder1983}. In the interaction between the star and its gravitational wake, energy and momentum are exchanged, accelerating some stars, decelerating others.

Chandrasekhar's approach has proved exceedingly successful despite the number of simplifying assumptions inherent in the model which are not strictly applicable to systems such as the Galactic core. We will not attempt to fix these deficiencies; the only modification is to substitute the velocity distribution.

Other authors have built upon the work of Chandrasekhar by considering inhomogeneous stellar distributions, via perturbation theory \citep{Lynden-Bell1972,Tremaine1984,Weinberg1986}, modelling energy transfer as anomalous dispersion, which adds higher order moments to the transfer probability \citep{Bar-Or2012}, or using the tools of linear response theory and the fluctuation-dissipation theory \citep[chapter 7]{Landau1958}, which allows relaxation of certain assumptions, such as homogeneity \citep{Bekenstein1992,Maoz1993,Nelson1999}. We will not attempt to employ such sophisticated techniques at this stage.

\subsection{Chandrasekhar's change in energy}\label{sec:Chandra}

We consider the interaction of a field star, denoted by 1, with a test star, 2; the change in energy squared from interaction over time $\delta t$ is approximately \citep[chapter 2]{Chandrasekhar1960}\footnote{As stressed by \citet{Antonini2011}, it is important to include both the piece for $v_1 \leq v_2$ and $v_1 \geq v_2$ when calculating the effects of dynamical friction.}
\begin{equation}
\Delta E^2(v_1) \simeq \dfrac{8\pi}{3} n(v_1)G^2m_1^2 m_2^2\ln\left(qv_2^2\right)\left\{\begin{array}{lr}\dfrac{v_1^2}{v_2}\vspace{1.0mm} & v_1 \leq v_2\\ \dfrac{v_2^2}{v_1} & v_1 \geq v_2 \end{array}\right\}\,\dd v_1\delta t.
\end{equation}
Here $v_1$ and $v_2$ are the initial velocities, and $m_1$ and $m_2$ are the masses; $n(v_1)$ is the number of stars per velocity element $\dd v_1$ which is calculated assuming that the density of stars is uniform.\footnote{The error introduced by this assumption can be partially absorbed by the appropriate choice of the Coulomb logarithm, which shall be introduced later \citep{Just2011}.} The logarithmic term includes
\begin{equation}
q = \dfrac{D_0}{G\left(m_1+m_2\right)},
\end{equation}
where $D_0$ is the maximum impact parameter \citep{Weinberg1986}. To eliminate the dependence upon $v_1$ requires a specific form for the velocity distribution.

\subsection{Velocity distributions}

The velocity space DF can be obtained by integrating out the spatial dependence in the full DF. As we are restricting our attention to the core and assuming spherical symmetry
\begin{equation}
f(v) = 4\pi\intd{0}{r\sub{c}}{r^2f(\mathcal{E})}{r},
\end{equation}
where $r\sub{c}$ is defined by \eqnref{r_c}.

The DF for unbound stars is assumed to be Maxwellian as in \eqnref{Unbound_DF}. We assume violent relaxation such that $\sigma_M = \sigma$. Performing the integral
\begin{align}
f_{\mathrm{u},\,M}(v) = {} & \dfrac{n_\star}{\left(2\pi\sigma^2\right)^{3/2}}C_M\epsilon\left(\dfrac{v^2}{2\sigma^2}\right),
\end{align}
introducing
\begin{equation}
\epsilon(w) = \recip{2}\left\{\exp(-w)\left[4\exp(1) + \Ei(w) - \Ei(1)\right] - \dfrac{2 + w + w^2}{w^3}\right\},
\end{equation}
where $\Ei(x)$ is the exponential integral.

The DF for bound stars is approximated as a simple power law as in \eqnref{Bound_DF}. The integral gives
\begin{equation}
f_{\mathrm{b},\,M}(v) = \dfrac{n_\star}{\left(2\pi\sigma^2\right)^{3/2}}k_M \left(\dfrac{v^2}{2\sigma^2}\right)^{p_M - 3}\begin{cases}
3 \Beta\left(\dfrac{v^2}{2\sigma^2}; 3 - p_M, 1 + p_M\right) & \dfrac{v^2}{2\sigma^2} \leq 1 \\
3 \Beta\left(3 - p_M, 1 + p_M\right) & \dfrac{v^2}{2\sigma^2} \geq 1
\end{cases},
\end{equation}
where $\Beta(x;a,b)$ is the incomplete beta function \citep[section 8.17]{Olver2010} and $\Beta(a,b) \equiv \Beta(1,a,b)$ is the complete beta function.

The velocity space density is related to the DF by
\begin{equation}
\dfrac{4\pi r\sub{c}^3}{3}n_M(v_1) = 4\pi v_1^2\left[f_{\mathrm{u},\,M}(v_1) + f_{\mathrm{b},\,M}(v_1)\right].
\end{equation}

\subsection{Defining the relaxation time-scale}

Using the specific forms for the velocity space density, we can calculate $\Delta E^2$. The functional form depends upon the velocity of the test star. If $v_2^2/2\sigma^2 < 1$, then
\begin{align}
\Delta E^2 \simeq {} & \dfrac{16}{3}\sqrt{2\pi}\dfrac{G^2m_1^2 m_2^2n_\star}{\sigma^3}\ln\left(qv_2^2\right) \left(\dfrac{v_2^2}{2\sigma^2}\right) \left[k \dfrac{3}{(2 - p)(1 + p)} {_3F_2}\left(-1-p,2-p,\dfrac{3}{2};3-p,\dfrac{5}{2};\dfrac{v_2^2}{2\sigma^2}\right) + C\right]\,\delta t,
\end{align}
where ${_3F_2}(a_1,a_2,a_3;b_1,b_2;x)$ is a generalised hypergeometric function \citep[section 16]{Olver2010}.\footnote{We have suppressed subscript $M$ for brevity.} The contribution from bound and unbound stars can be identified by the coefficients $k$ and $C$ respectively. It is necessary to sum over all the species to get the total value.

If $v_2^2/2\sigma^2 > 1$,
\begin{equation}
\Delta E^2 \simeq \dfrac{16}{3}\sqrt{2\pi}G^2m_1^2 m_2^2n_\star\sigma\ln\left(qv_2^2\right) \left(\dfrac{v_2^2}{2\sigma^2}\right)^{-1/2} \left[k\beta\left(\dfrac{v_2^2}{2\sigma^2};p\right) + C\alpha\left(\dfrac{v_2^2}{2\sigma^2}\right)\right]\,\delta t,
\end{equation}
where
\begin{align}
\alpha(w) = {} & \recip{2}\left\{3w^{-1/2} + 5 + \left[4\exp(1) - \Ei(1) + \Ei(w)\right]\left[\dfrac{3\sqrt{\pi}}{4}\erf\left(w^{1/2}\right) - \dfrac{3}{2}w^{1/2}\exp(-w)\right] - 3\sqrt{\pi}\exp(1)\erf(1) \right. \nonumber\\*
 & + \left. 3\left[{_2F_2}\left(\dfrac{1}{2},1;\dfrac{3}{2},\dfrac{3}{2};1\right) - w^{1/2}{_2F_2}\left(\dfrac{1}{2},1;\dfrac{3}{2},\dfrac{3}{2};w\right)\right]\right\}; \\
\beta(w;p) = {} & \begin{cases} \dfrac{3}{1/2 - p}\left[\Beta\left(\dfrac{5}{2},1+p\right) - \dfrac{3w^{p-1/2}}{2(2-p)}\Beta\left(3-p,1+p\right)\right] & p < \recip{2} \\
\dfrac{\pi}{32}\left[12 \ln(2) - 1 + 6 \ln(w)\right] & p = \recip{2} \end{cases} . 
\end{align}
Here ${_2F_2}(a_1,a_2;b_1,b_2;x)$ is another generalised hypergeometric function which originates from the integral
\begin{equation}
\intd{}{w}{\dfrac{\exp(w')\erf\left({w'}^{1/2}\right)}{w'}}{w'} = \dfrac{4w^{1/2}}{\sqrt{\pi}}{_2F_2}\left(\dfrac{1}{2},1;\dfrac{3}{2},\dfrac{3}{2};w\right).
\end{equation}

Combining the two regimes for $v^2/2\sigma^2$, we can simplify using approximate forms. For the bound contribution 
\begin{align}
\Delta E\sub{b}^2 \approx {} & 16\sqrt{2\pi}G^2m_1^2m_2^2n_\star\sigma\ln\left(qv_2^2\right) k \gamma\left(\dfrac{v_2^2}{2\sigma^2};p\right)\,\delta t,
\label{eq:Bound-approx}
\end{align}
where
\begin{align}
\gamma(w;p) = {} & \left(1 + w^4\right)^{-1}\left\{\left[\dfrac{3}{(1 + p)(2 - p)}w - \dfrac{9}{5(3-p)}w^2 + \dfrac{9p}{14(7-p)}w^3 \right] + w^{7/2}\beta\left(w;p\right)\right\}.
\end{align}
The resulting error, ignoring variation from $\ln\left(qv_2^2\right)$, is less than $3\%$.

The unbound contribution is
\begin{equation}
\Delta E\sub{u}^2 \approx \dfrac{16}{3}\sqrt{2\pi}G^2m_1^2m_2^2n_\star\sigma\ln\left(qv_2^2\right) C \Xi \dfrac{v_2^2}{2\sigma^2} \left[\Xi^2 + \left(\dfrac{v_2^2}{2\sigma^2}\right)^3\right]^{-1/2}\,\delta t,
\label{eq:Unbound-approx}
\end{equation}
where
\begin{equation}
\Xi = \lim_{w \rightarrow \infty}\left\{\alpha(w)\right\} \simeq 4.31.
\end{equation}
This reproduces the full function to better than $5\%$, ignoring variation from $\ln\left(qv_2^2\right)$.

The relaxation time-scale is the time interval $\delta t$ over which the squared change in energy becomes equal to the kinetic energy of the test star squared \citep{Bar-Or2012}
\begin{align}
\tau\sub{R} = {} & \left(\dfrac{m_2v_2^2}{2}\right)^2\dfrac{\delta t}{\Delta E^2} \\
 \approx {} & \dfrac{3v_2^4}{16\sqrt{2\pi}G^2n_\star\sigma\ln\left(qv_2^2\right)} \left(\sum_M M^2 \left\{k_M \gamma\left(\dfrac{v_2^2}{2\sigma^2};p_M\right) + C_M\Xi\left(\dfrac{v_2^2}{2\sigma^2}\right)\left[\Xi^2 + \left(\dfrac{v_2^2}{2\sigma^2}\right)^3\right]^{-1/2}\right\}\right)^{-1}.
\label{eq:tau_R1}
\end{align}

\subsection{Averaged time-scale}

The relaxation time-scale \eqnref{tau_R1} is for a particular velocity $v_2$. This is not of much use to describe the core or even a (non-circular) orbit where there is a velocity range. It is necessary to calculate an average. Both the change in energy squared and the kinetic energy are averaged. We use two averages: over the distribution of bound velocities to give the relaxation time-scale for the system and over a single orbit. The former is of use when considering the inner cut-off of stars due to collisions, and the latter when considering the transition to GW inspiral.

\subsubsection{System relaxation time-scale}\label{sec:system-ave}

The total number of bound stars in the core is
\begin{equation}
N_{\mathrm{b},\,M} = \dfrac{3}{3/2 - p_M}\dfrac{\Gamma(p_M + 1)}{\Gamma(p_M + 7/2)}N_\star k_M,
\end{equation}
where $\Gamma(x)$ is the gamma function. Using this as a normalisation constant, the probability of a bound star having a velocity in the range $v \rightarrow v + \dd v$ is
\begin{align}
4\pi v^2 p_{\mathrm{b},\,M}(v) \,\dd v = {} & \sqrt{\dfrac{2}{\pi}} \dfrac{v^2}{\sigma^3} \dfrac{\left(3/2 - p_M\right)\Gamma(p_M + 7/2)}{\Gamma(p_M + 1)} \left(\dfrac{v^2}{2\sigma^2}\right)^{p_M - 3}\left\{\begin{array}{lr}
\Beta\left(\dfrac{v^2}{2\sigma^2}; 3 - p_M, 1 + p_M\right) & \dfrac{v^2}{2\sigma^2} \leq 1 \\
\Beta\left(3 - p_M, 1 + p_M\right) & \dfrac{v^2}{2\sigma^2} \geq 1\end{array}\right\}\,\dd v.
\end{align}
The mean square velocity for bound stars in the core is then
\begin{align}
\overline{v^2_{M}} = {} & 3\sigma^2\dfrac{3/2 - p_M}{1/2 - p_M},
\end{align}
assuming $p_M < 1/2$.

In the case $p_M = 1/2$ we encounter a logarithmic divergence. This reflects there being a physical cut-off.\footnote{A similar diverge necessitates the introduction of $D_0$ in \apref{Chandra}.} We use $v\sub{max} = c/2$, which is the maximum speed reached on a bound orbit about a Schwarzschild BH. Marginally higher speeds can be reached for prograde orbits about a Kerr BH, but the maximal velocity for retrograde orbits is marginally lower. In reality, we expect the maximum velocity to be lower due to a depletion of orbits. We also suspect that a simple Newtonian description of these orbits is imprecise, but a full relativistic description is beyond the scope of this analysis. For $p_M = 1/2$,
\begin{align}
\overline{v^2_{M}} = {} & \dfrac{\sigma^2}{2}\left[12\ln(2) - 5 + 6 \ln\left(\dfrac{v\sub{max}^2}{2\sigma^2}\right)\right].
\end{align}
Using a typical value of $\sigma = 10^5\units{m\,s^{-1}}$,
\begin{equation}
\overline{v^2_{M}} \simeq 43\sigma^2.
\end{equation}
The mean square velocity is an order of magnitude greater than that for a Maxwellian distribution.

For the average of $\Delta E^2$, we replace $\ln\left(qv_2^2\right)$ by a suitable average, so it may be moved outside the integral \citep[chapter 2]{Chandrasekhar1960}. We replace it by the Coulomb logarithm \citep{Bahcall1976}
\begin{equation}
\ln\left(q\overline{v_2^2}\right) = \ln \Lambda_M \simeq \ln\left(\dfrac{M_\bullet}{M}\right).
\end{equation}
\citet{Just2011} find an extremely similar result fitting a Bahcall--Wolf cusp self-consistently. We calculate the averages for the bound and unbound populations individually and then combine these to obtain the total change for each species. We must distinguish between the bound population of field stars and the distribution of test stars over which we are averaging. We use subscripts $M$ and $M'$ respectively.\footnote{In a slight abuse of notation, we use for masses $m_M \equiv M$ and $m_{M'} \equiv M'$, and hope that it is clear that the summation is over the species.} The bound average may be approximated to about $10\%$ accuracy as
\begin{align}
\overline{\Delta E^2_{\mathrm{b},\,M'}} \approx {} & \sum_M\dfrac{2^{11/2}}{3}G^2M^2{M'}^2n_\star\sigma\ln\left(\Lambda_{M'}\right) k_M \dfrac{(3/2 - p_{M'})\Gamma(p_{M'} + 7/2)}{\Gamma(p_{M'} + 1)} \left[ \varpi\left(p_M,p_{M'}\right) + \iota \left(p_M,p_{M'}\right) \right] \delta t,
\end{align}
introducing
\begin{align}
\varpi\left(p_M,p_{M'}\right) = {} & \dfrac{30 + 36p_M + 25p_M^2 - p_{M'}\left(13 + 15p_M + 7 p_M^2\right) + p_{M'}^2\left(6 + 9p_M + 8p_M^2\right)}{210}; \\
\iota\left(p_M,p_{M'}\right) = {} & \Beta\left(3-p_{M'},1+p_{M'}\right) \begin{cases} \dfrac{3}{1/2 - p_M}\left[\dfrac{\Beta\left(5/2,1+p_M\right)}{2-p_{M'}} - \dfrac{3\Beta\left(3-p_M,1+p_M\right)}{2\left(2-p_M\right)\left(5/2 - p_M - p_{M'}\right)}\right] & p_M < \recip{2} \\
\dfrac{\pi}{32}\dfrac{4 + p_{M'} + 12 \left(2 - p_{M'}\right) \ln(2)}{\left(2-p_{M'}\right)^2} & p_M = \recip{2} \end{cases}.
\end{align}

The unbound component is approximately
\begin{align}
\overline{\Delta E^2_{\mathrm{u},\,M'}} \approx {} & \sum_M\dfrac{2^{11/2}}{3}G^2M^2{M'}^2n_\star\sigma\ln\left(\Lambda_{M'}\right) C_M \dfrac{(3/2 - p_{M'})\Gamma(p_{M'} + 7/2)}{\Gamma(p_{M'} + 1)} \nonumber \\
 & \times \left[\nu\left(p_{M'}\right) + \Xi\dfrac{\Beta\left(3-p_{M'},1+p_{M'}\right)}{2-p_{M'}}{_2F_1}\left(\recip{2},\dfrac{2-p_{M'}}{3};\dfrac{5-p_{M'}}{3};-\Xi^2\right) \right] \delta t,
\end{align}
where
\begin{equation}
\nu(p) = \begin{cases} \recip{1/2 - p}\left[\Beta\left(\dfrac{5}{2},1+p\right) - \Beta\left(3-p,1+p\right)\right] & p < \recip{2} \\
\dfrac{\pi}{96}\left[12 \ln(2) - 5\right] & p = \recip{2}
\end{cases} \; ,
\end{equation}
and we have used ${_2F_1}(a_1,a_2,;b_1;x)$, another hypergeometric function \citep[15.6.1]{Olver2010}. For consistency with the bound case, we have continued to use subscript $M'$. 

The total relaxation time for a species is
\begin{align}
\overline{\tau_{\mathrm{R,}\,M'}} = {} & \left(\dfrac{{M'}\overline{v_{M'}^2}}{2}\right)^2\dfrac{\delta t}{\overline{\Delta E^2_{\mathrm{b},\,M'}} + \overline{\Delta E^2_{\mathrm{u},\,M'}}} \\
 \approx {} & \dfrac{3}{2^{15/2}}\dfrac{\Gamma(p_{M'} + 1)}{(3/2 - p_{M'})\Gamma(p_{M'} + 7/2)}\dfrac{\overline{v_{M'}^2}^2}{G^2n_\star\sigma\ln\left(\Lambda_{M'}\right)} \left\{\sum_M k_M M^2 \left[ \varpi\left(p_M,p_{M'}\right) + \iota \left(p_M,p_{M'}\right)\right] \right. \nonumber \\*
 & + \left. \vphantom{ \left[ \dfrac{p_{M'}^2\left(6 + 9p_M + 8p_M^2\right)}{210}\right]} C_M M^2 \left[\nu\left(p_{M'}\right) + \Xi\dfrac{\Beta\left(3-p_{M'},1+p_{M'}\right)}{2-p_{M'}}{_2F_1}\left(\recip{2},\dfrac{2-p_{M'}}{3};\dfrac{5-p_{M'}}{3};-\Xi^2\right)\right]\right\}^{-1}.
\end{align}
Combining these to form an average for the entire system gives
\begin{equation}
\overline{\tau_{\mathrm{R}}} = \dfrac{\sum_{M'}N_{\mathrm{b,}\,M'}\overline{\tau_{\mathrm{R,}\,M'}}}{\sum_{M}N_{\mathrm{b,}\,M}}.
\label{eq:system-relax}
\end{equation}
The relaxation time-scale for individual components is used in determining the collisional cut-off as described in \secref{Collision}.

\subsubsection{Orbital average}\label{sec:orbital-ave}

We calculate the time-scale for an orbit, parametrized by $e$ and $r\sub{p}$, by averaging over one period.\footnote{We only consider bound orbits. The orbital relaxation time-scale is compared against the GW time-scale; the evolution of unbound orbits due to GW emission is negligible.} The mean square velocity is
\begin{equation}
\left\langle v^2\left(e,r\sub{p}\right)\right\rangle = \dfrac{GM_\bullet(1 - e)}{r\sub{p}}.
\end{equation}
The orbital average is calculated according to \citep[section 2.2b]{Spitzer1987}
\begin{equation}
\left\langle X\right\rangle = \recip{T}\intd{0}{T}{X(t)}{t},
\end{equation}
where $T$ is the orbital period. Despite our best efforts, we have been unsuccessful in obtaining analytic forms for the averaged changes in energy squared. Therefore, we compute them numerically. Switching to the orbital phase angle $\vartheta$, we define
\begin{align}
I\sub{b}(e,\varrho,p) = {} & \intd{0}{\pi}{\recip{(1 + e \cos\vartheta)^2}\gamma\left(\dfrac{1}{2(1+e)\varrho}\left(1+e^2+2e\cos\vartheta\right);p\right)}{\vartheta} \\
I\sub{u}(e,\varrho,\Xi) = {} & \intd{0}{\pi}{\dfrac{\Xi}{(1 + e \cos\vartheta)^2}\left[\dfrac{1}{2(1+e)\varrho}\left(1+e^2+2e\cos\vartheta\right)\right]\left\{\Xi^2 + \left[\dfrac{1}{2(1+e)\varrho}\left(1+e^2+2e\cos\vartheta\right)\right]^3\right\}^{-1/2}}{\vartheta}.
\end{align}
The orbital relaxation time-scale is then
\begin{align}
\left\langle\tau_{\mathrm{R},\,M'}\left(e,r\sub{p}\right)\right\rangle = {} & \left(\dfrac{GM_\bullet(1 - e)M'}{2r\sub{p}}\right)^2\dfrac{\delta t}{\left\langle\Delta E^2_{\mathrm{b},\,M'}\right\rangle + \left\langle\Delta E^2_{\mathrm{u},\,M'}\right\rangle} \\
 \approx {} & \dfrac{3}{64}\sqrt{\dfrac{\pi}{2}} \dfrac{M_\bullet^2(1 - e)^{1/2}}{n_\star \sigma r\sub{p}^2(1 + e)^{3/2}\ln\left(\Lambda_{M'}\right)} \left\{\sum_M \left[ k_M M^2 I\sub{b}\left(e,\dfrac{r\sub{p}}{r\sub{c}},p_M\right) + C_M M^2 I\sub{u}\left(e,\dfrac{r\sub{p}}{r\sub{c}},\Xi\right)\right]\right\}^{-1}.
\label{eq:orbital-relax}
\end{align}
This time-scale is defined similarly to the inspiral time-scale \eqnref{tGW-def}.

Diffusion in angular momentum proceeds over a shorter time, as defined by \eqnref{J-time}. Combining this with \eqnref{orbital-relax} gives the orbital angular momentum relaxation time-scale.

\subsection{Discussion of applicability}

In deriving the relaxation time-scales it has been necessary to make a number of approximations, both mathematical and physical. We have been careful to ensure that the mathematical inaccuracies introduced are of the order of a few percent, and subdominant to the errors inherent from the physical assumptions and uncertainties in astronomical quantities. There are two key physical approximations that may limit the validity of the results.

First, it was assumed that the density of stars is uniform. This is a pragmatic assumption necessary to perform integrals over the impact parameter and angular orientation. This is not the case; however, as a star travels on its orbit, it moves through regions of different densities, sampling a range of different density--impact parameter distributions. Since we are only concerned with averaged time-scales, this partially smears out changes in density \citep[cf.][]{Just2011}. To incorporate the complexity of the proper density distribution would greatly obfuscate the analysis.

Second, we have only considered transfer of angular momentum based upon the diffusion of energy, and not through resonant relaxation which enhances (both scalar and vector) angular momentum diffusion \citep{Rauch1996,Rauch1998,Gurkan2007,Eilon2009,Madigan2011}. This occurs in systems where the radial and azimuthal frequencies are commensurate. Orbits precess slowly leading to large torques between the orbits. These torques cause the angular momentum to change linearly with time over a coherence time-scale set by the drift in orbits. Over longer time periods, the change in angular momentum again proceeds as a random walk, increasing with the square root of time, as for non-resonant relaxation, but is still enhanced because of the change in the basic step size. Diffusion of energy remains unchanged; there could be several orders of magnitude difference in the two relaxation time-scales.

Resonant relaxation is important in systems with (nearly) Keplerian potentials, but is quenched when relativistic precession becomes significant: inside the Schwarzschild barrier \citep{Merritt2011}. It is less likely to be of concern for the orbits influenced by GW emission \citep{Sigurdsson1997} and should not be significant for our purposes.

The optimal approach would be to perform a full $N$-body simulation of the Galactic core. This would dispense with all the complications of considering relaxation time-scales and estimates for cut-off radii. Unfortunately, such a task still remains computationally challenging at the present time \citep[e.g.,][]{Li2012}.

\subsection{Time-scales for the Galactic core}\label{sec:tauGC}

Evaluating $\overline{\tau\sub{R}}$ for the Galactic core (\secref{GC-Param}) and comparing with $\tau\sub{R}\super{Max}$, \eqnref{tauMaxwell} using $\kappa = 0.34$, shows a broad consistency:
\begin{equation}
\overline{\tau\sub{R}} \simeq 2.0 \tau\sub{R}\super{Max}.
\end{equation}
This is reassuring since the standard Maxwellian approximation has been successful in characterising the properties of the Galactic core. We calculated $\tau\sub{R}\super{Max}$ for the dominant stellar component alone, which gives $\tau\sub{R}\super{Max}\simeq 4.5 \times 10^9\units{yr}$.

Looking at the time-scales for each species in turn:
\begin{equation}
\overline{\tau\sub{R,\,MS}} \simeq 1.7 \tau\sub{R}\super{Max};\quad \overline{\tau\sub{R,\,WD}} \simeq 1.6 \tau\sub{R}\super{Max};\quad \overline{\tau\sub{R,\,NS}} \simeq 2.1 \tau\sub{R}\super{Max}.
\end{equation}
Again there is good agreement.\footnote{\citet*{Freitag2006} found that using a consistent velocity distribution for the population of stars (from an $\eta$-model), instead of relying on the Maxwellian approximation, made negligible change to the dynamical friction time-scale. They did not consider a cusp as severe as $p = 0.5$.} For BHs,
\begin{equation}
\overline{\tau\sub{R,\,BH}} \simeq 48 \tau\sub{R}\super{Max}.
\end{equation}

The time-scales for the lighter components are of the order of the Hubble time; the BH time-scale is much longer on account of the higher mean square velocity. This may indicate that the BH population is not fully relaxed \citep[cf.][]{Antonini2011}: there has not been sufficient time for objects to diffuse on to the most tightly bound orbits (in which case, the mean square velocity would be lower). We expect that many of the most tightly bound BHs are not in a relaxed state, since GW inspiral is the dominant effect in determining the profile. This would deplete some of the innermost orbits and lower the mean square velocity for the population. Since we do not consider the collisional disruption of BHs, we do not use $\overline{\tau\sub{R,\,BH}}$ in our model; it therefore has no influence on our results.

The long BH time-scale also inevitably includes an artifact of our approximation that the system is homogeneous: in reality the BHs, being more tightly clustered towards the centre, pass through regions with greater density (both because of a higher number density and a greater average object mass). Therefore, we expect the true relaxation time-scale to be reduced. 

Formation of the cusp can occur over shorter time than the relaxation time-scale \citep{Bar-Or2012}. It should proceed on a dynamical friction time-scale $\tau\sub{DF} \approx (M_\star/M')\overline{\tau_{\mathrm{R},\,M'}}$ \citep[section 3.4]{Spitzer1987}. This reduces the difference between the different species, but does not make it obvious that the cusp has had sufficient time to form, especially if there has been a merger in the Galaxy's history which disrupted the central distribution of stars \citep{Gualandris2012}. Fortunately, observations of the thick disc indicate that there has not been a major merger in the last $10^{10}\units{yr}$ \citep{Wyse2008}. Minor mergers, where (globular) clusters spiral in towards the MBH, have been suggested as a means of building the stellar population that is consistent with current observations \citep{Antonini2011a,Antonini2013}. These could prevent the cusp from forming if there has not been sufficient time for the stars to relax post-merger. In any case, the time taken to form a cusp depends upon the initial configuration of stars, and so depends upon the Galaxy's history. 

The existence of a cusp is a subject of debate. \citet{Preto2010} conducted $N$-body simulations to investigate the effects of strong mass segregation \citep{Alexander2009, Keshet2009} and found that cusps formed in a fraction of a (Maxwellian) relaxation time \citep{Amaro-Seoane2011}. \citet{Gualandris2012} conducted similar computations and found that cores are likely to persist for the dominant stellar popular; intriguingly, cusp formation amongst BHs is quicker, but still takes at least a (Maxwellian) relaxation time. We cannot add further evidence to settle the matter. Our state of understanding may be improved following the passage through periapse of the gas cloud G2 this year \citep{Bartos2013}. For definiteness, we have assumed that a cusp has formed in our calculations.

Time-scales for individual orbits range over many orders of magnitude. The longest are for the most tightly bound: the cusp forms from the outside-in, and these orbits may not yet be populated. The shortest time-scales are for the most weakly bound orbits, those with large periapses and eccentricities. The orbital period can be much shorter than these time-scales, highlighting the fringe where the Fokker--Planck approximation is not appropriate \citep{Spitzer1972}. The variation in the time-scale is exaggerated by neglecting the spatial variation in the stellar population.

When comparing GW inspiral time-scales and orbital angular momentum time-scales, equality can occur for times far exceeding the Hubble time. This only occurs for lower eccentricities, which are not of interest for bursts. However, it may be interesting to consider the stellar distribution in this region, which is not relaxed but dominated by GW inspiral. Since inspiral takes such a huge time to complete, it is possible that there is a pocket of objects currently mid-inspiral that reflect the unrelaxed distribution.

\section{Evolution of orbital parameters from gravitational wave emission}

\subsection{Bound orbits}

For bound orbits, we can define a GW inspiral time from the orbit-averaged change in the orbital parameters. Using the analysis of \citet{Peters1964} for Keplerian binaries, the averaged rates of change of the periapsis and eccentricity are
\begin{align}
\left\langle\diff{r\sub{p}}{t}\right\rangle = {} & -\dfrac{64}{5}\dfrac{G^3M_\bullet M(M_\bullet + M)}{c^5r\sub{p}^3}\dfrac{(1 - e)^{3/2}}{(1 + e)^{7/2}}\left(1 - \dfrac{7}{12}e + \dfrac{7}{8}e^2 + \dfrac{47}{192}e^3\right) \\
\left\langle\diff{e}{t}\right\rangle = {} & -\dfrac{304}{15}\dfrac{G^3M_\bullet M(M_\bullet + M)}{c^5r\sub{p}^4}\dfrac{e(1 - e)^{3/2}}{(1 + e)^{5/2}}\left(1 + \dfrac{121}{304}e^2\right).
\end{align}
For a circular orbit, the inspiral time from initial periapsis $r\sub{p0}$ is
\begin{equation}
\tau\sub{c}(r\sub{p0}) = \dfrac{5}{256}\dfrac{c^5r\sub{p0}^4}{G^3M_\bullet M(M_\bullet + M)}.
\end{equation}
For an orbit of non-zero eccentricity ($0 < e < 1$), we can solve for the periapsis as a function of eccentricity
\begin{equation}
r\sub{p}(e) = \mathcal{R}(1 + e)^{-1}\left(1 + \dfrac{121}{304}e^2\right)^{870/2299}e^{12/19},
\end{equation}
where $\mathcal{R}$ is fixed by the initial conditions: for an orbit with initial eccentricity $e_0$,
\begin{equation}
\mathcal{R}(e_0) = (1 + e_0)\left(1 + \dfrac{121}{304}e_0^2\right)^{-870/2299}e_0^{-12/19}r\sub{p0}.
\end{equation}
The inspiral is complete when the eccentricity has decayed to zero; the inspiral time is \citep{Peters1964}
\begin{equation}
\tau\sub{insp}(r\sub{p0},e_0) = \intd{0}{e_0}{\dfrac{15}{304}\dfrac{c^5\mathcal{R}^4}{G^3M_\bullet M(M_\bullet + M)}\dfrac{e^{29/19}}{(1-e^2)^{3/2}}\left(1 + \dfrac{121}{304}e^2\right)^{1181/2299}}{e}.
\end{equation}
This is best evaluated numerically, but it may be written in closed form as
\begin{equation}
\tau\sub{insp}(r\sub{p0},e_0) = \tau\sub{c}(r\sub{p0})(1 + e_0)^4\left(1 + \dfrac{121}{304}e_0^2\right)^{-3480/2299} F_1\left(\dfrac{24}{19};\dfrac{3}{2},-\dfrac{1181}{2299};\dfrac{43}{19};e_0^2,-\dfrac{121}{304}e_0^2\right),
\label{eq:Bound_inspiral}
\end{equation}
using the Appell hypergeometric function of the first kind $F_1(\alpha;\beta,\beta';\gamma;x,y)$ \citep[16.15.1]{Olver2010}.\footnote{For small eccentricities, $\tau\sub{insp}(r\sub{p0},e_0) \simeq \tau\sub{c}(r\sub{p0})[1 + 4e_0 + (273/43)e_0^2 + \order{e_0^3}]$.}

\subsection{Unbound orbits}\label{sec:Unbound}

Unbound objects only pass through periapsis once. We therefore expect the orbital change from gravitational radiation to be small. Following the approach of \citet{Turner1977}, we can calculate the evolution in the eccentricity and periapse of an unbound Keplerian binary. The change in fractional eccentricity over an orbit, approximating the orbital parameters as constant, is
\begin{equation}
\dfrac{\Delta e}{e} = -\dfrac{608}{15}\Sigma\left[\recip{(1+e)^{5/2}}\left(1 + \dfrac{121}{304}e^2\right)\cos^{-1}\left(-\recip{e}\right) + \dfrac{(e - 1)^{1/2}}{e^2(1+e)^2}\left(\dfrac{67}{456} + \dfrac{1069}{912}e^2 + \dfrac{3}{38}e^4\right)\right],
\end{equation}
introducing the dimensionless parameter
\begin{equation}
\Sigma = \dfrac{G^{5/2}M_\bullet M(M_\bullet+ M)}{c^5r\sub{p}^{5/2}}.
\end{equation}
Similarly, the fractional change in periapsis is
\begin{equation}
\dfrac{\Delta r\sub{p}}{r\sub{p}} = -\dfrac{128}{5}\Sigma\left[\recip{(1+e)^{7/2}}\left(1 - \dfrac{7}{12}e + \dfrac{7}{8}e^2 + \dfrac{47}{192}e^3\right)\cos^{-1}\left(-\recip{e}\right) - \dfrac{(e - 1)^{1/2}}{e(1 + e)^3}\left(\dfrac{67}{288} - \dfrac{13}{8}e + \dfrac{133}{576}e^2 - \dfrac{1}{4}e^3 - \dfrac{1}{8}e^4\right)\right].
\end{equation}
Both of these changes obtain their greatest magnitudes for large eccentricities, then
\begin{equation}
\dfrac{\Delta e}{e} \simeq \dfrac{\Delta r\sub{p}}{r\sub{p}} \simeq -\dfrac{16}{5}\Sigma e^{1/2}.
\end{equation}
For extreme mass-ratio binaries, as is the case here, the mass-ratio is a small quantity
\begin{equation}
\eta = \dfrac{M}{M_\bullet} \ll 1.
\end{equation}
The smallest possible periapsis is of the order of the Schwarzschild radius of the MBH, such that 
\begin{equation}
r\sub{p} = \alpha\dfrac{GM_\bullet}{c^2}; \quad \alpha > 1.
\end{equation}
These give
\begin{equation}
\Sigma = \dfrac{\eta}{\alpha^{5/2}} < \eta \ll 1.
\end{equation}
Hence, the changes in the orbital parameters become significant for
\begin{equation}
e \sim \dfrac{25}{256}\dfrac{\alpha^5}{\eta^2} > \dfrac{25}{256}\recip{\eta^2}.
\end{equation}
Such orbits should be exceedingly rare, and so it is safe to neglect inspiral for unbound orbits.

\end{onecolumn}

\bsp

\label{lastpage}

\end{document}